\documentclass{aa} % normal two columns
\usepackage{graphicx}
\usepackage{mathtools} % include amsmath
\usepackage[varg]{txfonts} % Postscript TX Times-fonts
\usepackage{xcolor}
\usepackage{xspace}
\usepackage{multirow}
\usepackage{rotating, lscape}
\usepackage{float}

\usepackage{hyperref}

\graphicspath{{./}{figs/}}

\def\J{\textit{J}\xspace}
\def\H{\textit{H}\xspace}
\def\Ks{\textit{$K_s$}\xspace}
\def\Av{$A_{\rm V}$\xspace}

\def\Rv{$R_{\rm V}$\xspace}

\def\NHth{NH$_3$\xspace}
\def\Nt{N$_2$\xspace}
\def\NtHp{N$_2$H$^+$\xspace}
\def\NtDp{N$_2$D$^+$\xspace}
\def\DCOp{DCO$^+$\xspace}

\def\Ht{H$_2$\xspace}

\def\Hthp{H$_3^+$\xspace}
\def\HtDp{H$_2$D$^+$\xspace}
\def\oHtDpgrd{1$_{10}$--1$_{11}$\xspace}
\def\DtHp{D$_2$H$^+$\xspace}
\def\pDtHpgrd{1$_{10}$--1$_{01}$\xspace}
\def\Dthp{D$_3^+$\xspace}
\def\CetO{C$^{18}$O\xspace}

\def\twvCO{$^{12}$CO\xspace}
\def \H{\textit{H}\xspace}
\def \B{\textit{B}}
\def \mJybm{mJy~beam$^{-1}$\xspace}

\def\arcmin{\mbox{$^{\prime}$}\xspace}
\def\arcsec{\mbox{$^{\prime\prime}$}\xspace}
\newcommand\micron{\mbox{$\mu$m}\xspace}

\begin{document}

\title{Deuterium fractionation of the starless core \object{L\,1498}}

\author{Sheng-Jun Lin\inst{1,2,3}
        \and Shih-Ping Lai\inst{2,3}
        \and Laurent Pagani\inst{4}
        \and Charl\`ene Lef\`evre\inst{5}
        \and Travis J. Thieme\inst{1}
        }

\institute{Academia Sinica Institute of Astronomy and Astrophysics (ASIAA), No. 1, Section 4, Roosevelt Road, Taipei 10617, Taiwan\\
            \email{shengjunlin@asiaa.sinica.edu.tw, slai@phys.nthu.edu.tw}
            \and
            Institute of Astronomy, National Tsing Hua University (NTHU), No.\,101, Section 2, Kuang-Fu Road, Hsinchu 30013, Taiwan
            \and
            Center for Informatics and Computation in Astronomy (CICA), NTHU, No. 101, Section 2, Kuang-Fu Road, Hsinchu 30013, Taiwan
            \and
            LERMA \& UMR8112 du CNRS, Observatoire de Paris, PSL  University, Sorbonne Universit\'es, CNRS, F-75014 Paris, France
            \and
            Institut de Radioastronomie Millim\'etrique (IRAM), 300 rue de la Piscine, 38400 Saint-Martin d’H\`eres, France
            }

\date{Received abc, dd, yyyy; accepted abc, dd, yyyy}

%Context
\abstract{Molecular deuteration is commonly seen in starless cores
and is expected to occur on a timescale comparable to that of the core contraction.
Thus, the deuteration serves as a chemical clock, allowing us to investigate dynamical theories of core formation.}
%Aims
{We aim to provide a 3D cloud description for the starless core L\,1498 located in the nearby low-mass star-forming region Taurus, and explore the possible core formation mechanism of L\,1498.}
%Methods
{We carried out non-local thermal equilibrium radiative transfer with multi-transition observations of
the high-density tracer \NtHp to derive the density and temperature profiles of the L\,1498 core. 
Combining with the spectral observations of the deuterated species, ortho-\HtDp, \NtDp, and \DCOp, we derived the abundance profiles for observed species and
performed chemical modeling of the deuteration profiles across L\,1498 
to constrain the contraction timescale. 
}
%Results
{We present the first ortho-\HtDp (\oHtDpgrd) detection toward L\,1498.
We find a peak molecular hydrogen density of 
$1.6_{-0.3}^{+3.0}\times10^{5}$~cm$^{-3}$, 
a temperature of 
7.5$_{-0.5}^{+0.7}$~K, and 
a \NtHp deuteration of 
0.27$_{-0.15}^{+0.12}$ in the center.
}
%Conclusions
{We derive a lower limit of the core age for L\,1498 of 0.16~Ma which is compatible with the typical free-fall time,
indicating that L\,1498 likely formed rapidly.
}


\maketitle

\section{Introduction}
Deuterium fractionation is closely related to star formation.
Almost all deuterium was formed through primordial nucleosynthesis after the birth of the Universe,
and there is no other process significantly producing deuterium afterward \citep{Wannier80}. 
The atomic deuterium fraction (D/H) is measured as 1.6$\times$10$^{-5}$ in the Local Bubble \citep{Linsky07}. 
In dark clouds, hydrogen primarily exists in its molecular form. HD serves as the primary deuterium reservoir, inheriting the deuterium fraction to be $3.2\times10^{-5}$. 
However,
deuterium fractionation in other hydrogen-containing molecules is found to be enhanced by several orders of magnitude in star-forming regions, especially, in starless cores \citep{Ceccarelli14}. 

Starless cores are the potential birthplaces of future stars and planets, where the key species leading the deuterium fractionation is trihydrogen cation, \Hthp. 
The deuteration is in fact an exothermic reaction owing to the larger mass of deuterium and thus the lower zero-point vibrational energies (ZPVEs) of the deuterated isotopologues \citep[e.g.,][]{Hugo09}.
Consequently, the low-temperature environment in starless cores (typically $\lesssim$ 10~K) favors the deuterium fractionation of \Hthp, where \Hthp repeatedly reacts with HD to form \HtDp, \DtHp, and \Dthp sequentially. 
On the other hand, the depletion of heavy-element-bearing species, including CO (which is a destruction partner of \Hthp), from the gas phase in starless cores makes \Hthp isotopologues (\Hthp, \HtDp, \DtHp, and \Dthp) become relatively abundant. 
At the center of starless cores, chemical modeling has shown that the \Hthp isotopologues are indeed the most abundant molecular ions \citep{Walmsley04, Flower05, Pagani09b}.
In the gas phase, deuterated trihydrogen cations can easily react with other molecules (e.g., \Nt and CO) to transfer the deuterium to form deuterated molecules (e.g., \NtDp and \DCOp) via neutral-ion reactions. 
Therefore, the deuterium enrichment developed in starless cores is closely tied to the deuterium fractionation of \Hthp. 

Although starless cores are natural deuteration factories, the deuteration of \Hthp does not freely proceed but is limited by the presence of ortho-\Ht (o-\Ht) because o-\Ht has a higher ZPVE of 170~K compared with para-\Ht, which helps to overcome the energy barrier in opposite direction (i.e., rehydrogenation), competing against the deuteration \citep{Pineau91}. 
It has been shown that the ZPVE differences among the different nuclear spin states of \Hthp isotopologues also contribute to lowering the energy barrier in rehydrogenation \citep{Pagani92a, Flower04b, Hugo09}.
The \Ht molecule is thought to be produced on the grain surface with its statistical equilibrium ortho-to-para ratio of \Ht (OPR(\Ht)) of 3 and expected to slowly decrease to $\ll$10$^{-2}$ within $\sim$10~Ma \citep{Pagani13}. 
The initial high OPR(\Ht) (i.e., abundant o-\Ht) is therefore a bottleneck in the deuterium fractionation.

One of the important questions in low-mass star formation is whether starless cores follow fast collapse or slow collapse models.
The former makes cores form through gravo-turbulent fragmentation within a few free-fall times, which is typically less than $\sim$1~Ma \citep{MacLow04, Padoan04, Ballesteros-Paredes07, Hennebelle12, Hopkins12, Federrath12}. The latter 
represents slowed-down collapse due to the support from magnetic fields against gravity, where
the collapse process could become about ten times longer than the former \citep{Shu87, Tassis04, Mouschovias06}.
The above core-collapse scenarios assume that the parent clouds are quasi-stationary while isolated cores are collapsing due to the lack of sufficient support.
While our focus in this paper is on the low-mass star-forming region, we also note that low-mass cores can also result from clump-fed scenarios associated with massive star formation. These scenarios include competitive accretion \citep{Bonnell01, Bonnell04}, global hierarchical collapse \citep{Vazquez-Semadeni17, Vazquez-Semadeni19}, and inertial flow \citep{Padoan20} models. The core formation could be more dynamical because the parent clouds/clumps/filaments are not quasi-stationary but involve multi-scale and anisotropic collapses.

Since the timescale of the aforementioned decreasing behavior of OPR(\Ht) is comparable to the timescale of core contraction,
the OPR(\Ht) effectively serves as an ideal chemical clock tracing the core contraction.
However, due to the absence of observable \Ht (sub)mm-transitions, deuterium fractionation emerges as an excellent proxy for the OPR(\Ht).
Consequently, the deuterium fractionation in the cores provides an accessible chemical clock. This allows for the estimation of the contraction timescale, thereby enabling us to differentiate between two distinct dynamical models of core formation \citep{Flower06a, Pagani09b, Pagani13, Kong15, Kong16, Kortgen17, Kortgen18, Bovino19, Bovino21}.

Owing to the depletion effect, the heavy species (e.g., CO, CS) are mostly removed from the gaseous portion of starless cores,
making these species ineffective to trace $n_{\rm H_2}$ and $T_{\rm kin}$.
On the other hand, light N-bearing species (e.g., \NtHp, \NHth) are relatively insensitive to the depletion effect and known to be dense-gas tracers.
Observationally, we often see that \NtHp and \NHth line emissions are spatially anti-correlated with CO line emission \citep{Bergin02, Tafalla02, Fontani06}, showing that they are confined in starless cores. 
In addition, \NtHp (and \NtDp) is solely formed in the gas phase, directly linked to the \Hthp deuterium fractionation, and 
\NtHp usually shows largest deuterium fractions (\NtDp/\NtHp) compared with the other N-bearing species \citep[e.g., \NHth, HCN, ][]{Pagani07, Ceccarelli14, Fontani15}. The \NtHp isotopologues are excellent high-density/deuteration tracers of starless cores.

To constrain the contraction timescale, 
\citet{Pagani07, Pagani09b, Pagani12} have 
developed an approach based on 
\begin{enumerate}
    \item the deuteration profile of \NtHp plus the abundance profile of ortho-\HtDp (o-\HtDp) across starless cores, and
    \item a deuterium chemical network that includes each spin-state of \Ht, \Hthp isotopologues and assumes a complete depletion condition in heavy species except for CO and \Nt.
\end{enumerate}
In this approach, the starless core is approximated with an onion-like physical model consisting of multiple shells. 
To evaluate chemical abundance, density, and temperature profiles at each shell, 
the nonlocal thermal equilibrium (non-LTE) hyperfine radiative transfer
and dust extinction measurements are performed 
with radio observations (multi-transition of \NtHp, \NtDp, \DCOp, and the o-\HtDp ground transition) and infrared observations, respectively.
In addition to the \NtDp/\NtHp ratio which traces the deuterium enrichment of the inner structure in the starless core,
the abundance of o-\HtDp is another key measurement to constrain the balance between the four \Hthp isotopologues 
(i.e., the different levels among the enhancements of \HtDp, \DtHp, \Dthp with respect to \Hthp).
Consequently, the contraction timescale can be estimated with time-dependent chemical modeling. 
In addition, the depletion factors of CO and \Nt can also be derived in the aspect of volume densities, instead of column densities, from the onion-like physical model.

With the above approach, it is found that
the starless core L\,183, located in Serpens, is presumably older than 0.15--0.2~Ma \citep{Pagani09b} and is consistent with the fast collapse model \citep{Pagani13}. 
We have also applied this approach to L\,1512, an isolated spherically symmetric starless core located in Auriga \citep{Lin20}.
We found that L\,1512 is consistent with the slow collapse model in contrast to L\,183 
because our time-dependent chemical analysis shows that L\,1512 is presumably older than 1.43~Ma, much larger than the typical free-fall time. 
In addition, the similarity between the \Nt and CO abundance profiles in L\,1512 suggests that
L\,1512 has probably been living long enough so that \Nt chemistry has reached a steady-state.
In this paper, we focus on a slightly asymmetric starless core, L\,1498.

L\,1498 is a nearby starless core embedded in a filament that is located at a distance of 140~pc in the Taurus molecular cloud \citep{Myers83a, Beichman86, Myers91, Tafalla02}.
The L\,1498 envelope has significant infall motion detected with the blue asymmetry feature in CS spectra \citep{Lee99b, Lee01, Lee11}, while the L\,1498 core is quiescent and shows a small nonthermal velocity dispersion of 0.054~km~s$^{-1}$ and an averaged gas temperature of 7.7~K derived from the line widths of HC$_3$N and \NHth \citep{Fuller93}.
The central density of L\,1498 was constrained to a range of $\sim0.1$--$1.35\times10^5$~cm$^{-3}$ from continuum observations but is sensitive to the different wavelengths, temperature profiles, and dust opacities adopted by different authors
\citep{Langer01,Shirley05,Tafalla02, Tafalla04,Magalhaes18}.
Recently, the ALMA-ACA 1~mm continuum survey, FREJA, found no substructures at the central 1000~au scale in L\,1498, suggesting a central density $\lesssim3\times10^5$~cm$^{-3}$
and/or a flat inner density structure \citep{Tokuda20}. 
With polarization observations,
\citet{Levin01} constrained an upper limit on the line-of-sight \B-field strength of 100~$\mu$G, 
while \citet{Kirk06} derived a plane-of-sky \B-field strength of 10$\pm$7~$\mu$G, suggesting that the L\,1498 core region is virialized and the thermal support might be superior to the magnetic support, implying a magnetically supercritical state.

\citet{Aikawa05} conducted a chemo-dynamical coupled modeling 
of a Bonner-Ebert sphere
and found that a nearly thermally supported model with the gravity-to-pressure ratio of 1.1 can generally reproduce the infall feature and depletion of CO and CS in L\,1498. 
\citet{Yin21} performed non-ideal magnetohydrodynamic (MHD) simulations with ambipolar diffusion,
coupled with a chemical network, focusing on the collapse of a static sphere of constant density. 
They compared synthetic \NtHp, CS, and \CetO line profiles with observations under two initial conditions, namely magnetically subcritical or supercritical clouds. 
Their findings also suggest a consistent conclusion that L\,1498 core gradually evolved from an initially subcritical condition to a supercritical condition at present. However, we note that their radiative line modeling did
not consider \NtHp hyperfine lines \citep[see Appendix C in ][]{Priestley22}
As a result, the authors  
emphasized qualitative aspects in the comparison between their synthetic spectra and observations 
in terms of linewidths in their approach.
The chemical differentiation nature in L\,1498 has been widely studied in previous works
\citep{Lemme95, Kuiper96, Wolkovitch97, Willacy98, Lai00, Tafalla02, Tafalla04, Young04, Tafalla06, Ford11}, suggesting that
many species (e.g., CO, CCS, CS) are subject to depletion in the core center and their emissions show ring-like or peanut-like (double-peaked) morphologies.
For the nitrogen-bearing species \NHth and \NtHp at the core center, 
\citet{Tafalla04} found that their abundances are enhanced by a factor of a few through a spectral fitting,
while chemical modeling studies reported that both abundance enhancement and depletion features are possible \citep{Aikawa03, Aikawa05, Holdship17}.
However, we note that \citet{Tafalla04} ignored the hyperfine overlaps in their spectral fitting and used their own hyperfine structure collisional coefficients based on educated guess since the actual ones were not yet available \citep{Daniel05, Lique15}.
Thus, a detailed non-LTE radiative transfer modeling is needed.

On the other hand, in addition to deuterium, 
different species have been used as chemical clocks but
yielded different chemical timescales.
\citet{Maret13} derived a timescale of $\sim$0.3~Ma by modeling the \CetO (1--0) and H$^{13}$CO$^+$ (1--0) emissions toward L\,1498 using a gas-phase chemical code with gas-grain interactions included. 
\citet{Jimenez-Serra21} derived a timescale of $\sim$0.09~Ma with their observations of complex organic molecules (COMs) using a three-phase (gas, grain ice surface, and grain ice bulk) chemical code. 
This discrepancy of a factor of $\sim$3 in the chemical timescales might be related to the different initial chemical abundance conditions and chemical reaction sets in the model and/or the different spatial region that molecules are tracing.
Moreover, these chemical models are pseudo-time dependent, which does not include the physical evolution of the core. The above chemical timescales are thus lower limits of the core age.
Investigating different chemical clocks could help to understand their robustness.

In this paper, we use multi-transition radio observations of the high-density tracer \NtHp and infrared observations to derive the density and temperature profiles in the L\,1498 core. 
Combining with the emission spectra of the deuterated species, o-\HtDp, \NtDp, and \DCOp, we aim to constrain the contraction timescale of L\,1498 and the possible
core formation models via deuterium chemical modeling.
We describe our observations in Sect.\,\ref{sec:observation_L1498} and present them in Sect.\,\ref{sec:results_L1498}.
In Sect.\,\ref{sec:analysis_L1498}, we perform the nonlocal thermal equilibrium (non-LTE) radiative transfer modeling and time-dependent chemical modeling.
In Sect.\,\ref{sec:discussion_L1498}, we discuss the possible formation mechanism of L\,1498. 
We summarize our results in Sect.\,\ref{sec:conclusion_L1498}.

\section{Observations and data reduction}\label{sec:observation_L1498}

\begin{table*}
\centering
    \caption{Observational parameters.}
    \begin{tabular}{lrcccc}
    \hline\hline
    Line & Frequency$^{\tablefootmark{a}}$ & $\delta$v  & $T_{\rm rms}$$^{\tablefootmark{b}}$ & $\theta_{\rm MB}$$^{\tablefootmark{c}}$ & $\eta_{\rm MB}$ \\ 
    & (MHz) & (m\,s$^{-1}$) & (mK)  & (\arcsec) & \\
    \hline
    \noalign{\smallskip}
    \multicolumn{6}{c}{Institut de Radio-Astronomie Millim\'{e}trique
(IRAM) 30-m}\\
    \noalign{\smallskip}
    \hline
    \noalign{\smallskip}
    \NtHp (1--0) &  93173.764&31& 22--72  &26&0.85\\
    \DCOp (2--1) & 144077.282&41& 28--53  &17&0.80\\
    \NtDp (2--1) & 154217.181&38& 25--42  &16&0.78\\
    \DCOp (3--2) & 216112.582&68& 48--104 &11&0.66\\
    \CetO (2--1) & 219560.358&67& 55--123 &11&0.66\\
    \NtHp (3--2) & 279511.832&42& 30--68  & 9&0.56\\
    \noalign{\smallskip}
    \hline
    \noalign{\smallskip}
    \multicolumn{6}{c}{Green Bank Telescope (GBT) 100-m}\\
    \noalign{\smallskip}
    \hline
    \noalign{\smallskip}
    \DCOp (1--0) & 72039.303&48& 55--98 & 11&0.51\\
    \NtDp (1--0) & 77109.616&44& 25--30 & 10&0.51\\
    \noalign{\smallskip}
    \hline
    \noalign{\smallskip}
    \multicolumn{6}{c}{James Clerk Maxwell Telescope (JCMT) 15-m}\\
    \noalign{\smallskip}
    \hline
    \noalign{\smallskip}
    \HtDp (\oHtDpgrd) & 372421.356&49& 30--52 &13&0.60\\
    \NtHp (4--3)      & 372672.526&49& 38--58 &13&0.60\\
    \noalign{\smallskip}
    \hline
    \end{tabular}
    \tablefoot{
    \tablefoottext{a}{\NtHp and \NtDp frequencies are taken from 
    \citet{Pagani09a}. The frequencies correspond to the strongest
    hyperfine component for each transition.}
    \tablefoottext{b}{The rms noise is expressed in the $T_{\rm A}^*$ scale \citep{Kutner81}.}
    \tablefoottext{c}{The beam size is the HPBW.}
    }
\label{tab:obs_L1498}
\end{table*}

\begin{figure}
	\centering
    \includegraphics[scale=0.34]{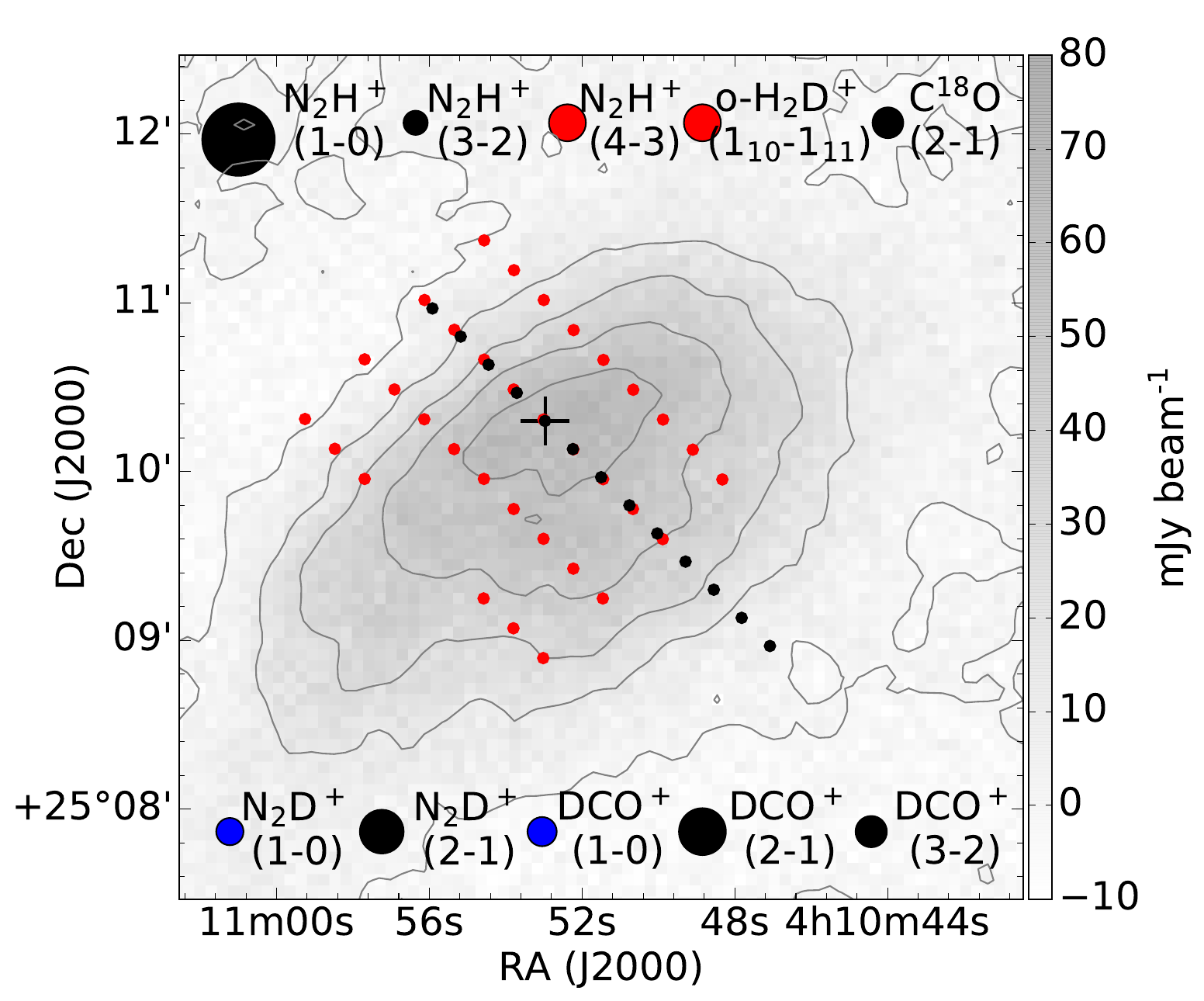}
	\caption{
    Multi-pointing grids overlaid with the SCUBA-2 850~\micron map.
    The black dots in a ($\Delta$RA, $\Delta$Dec)=(10\arcsec, 10\arcsec)-spacing 45$^\circ$-cut show the pointings of IRAM 30-m and GBT observations. 
    The red dotted grid shows the pointings of JCMT observations.
    The circles at the top and bottom indicate the beam sizes ($\theta_{\rm MB}$) of each spectral observation with the same color as the pointing grid except that GBT beam sizes are shown in blue.
    The 850~\micron map is shown in grayscale with a beam size of 14\arcsec and overlaid with its contours at 0\%, 20\%, 40\%, 60\%, and 80\% of its peak intensity at 78~\mJybm.
    }
    \label{fig:pointing_L1498}
\end{figure}

\subsection{Spectral observations}

We conducted radio and sub-millimeter (submm) observations toward L\,1498 using the 
Institut de Radio-Astronomie Millim\'{e}trique
(IRAM) 30-m telescope, 
the James Clerk Maxwell Telescope (JCMT),
and the Green Bank Telescope (GBT).
Observational parameters are summarized in Table\,\ref{tab:obs_L1498} and
the pointings of spectral line observations are shown in Fig.\,\ref{fig:pointing_L1498}.

\subsubsection{IRAM, JCMT, and GBT observations}

We observed L\,1498 in \NtHp (1--0), \NtHp (3--2), \NtDp (2--1), \DCOp (2--1), \DCOp (3--2), and \CetO (2--1) 
using the IRAM 30-m telescope in December 2013, May and October 2014, and September 2017.
The observations were performed in frequency-switching mode 
using the dual polarization Eight MIxer Receiver (EMIR), the VErsatile SPectral Autocorrelator (VESPA), and the Fourier Transform Spectrometer (FTS).
The spectra were observed on a ($\Delta$RA, $\Delta$Dec)=(10\arcsec, 10\arcsec)$_{\rm J2000}$-spacing grid in a northeast-southwest cut across the core center at (RA, Dec)$_{\rm J2000}$ = 
(4$^{h}$10$^{m}$52\fs97, +25\degr10\arcmin18\farcs0).
The data were subsequently folded and baseline subtracted with CLASS\footnote{http://www.iram.fr/IRAMFR/GILDAS}. 
In addition, we complemented our observations with the 
data of \NtHp (1--0) and \CetO (1--0) obtained by \citet{Tafalla04}. 
We refer readers to the observational details described by \citet{Tafalla04}.
Their observation was conducted on a ($\Delta$RA, $\Delta$Dec)$_{\rm B1950}$=(20\arcsec, 20\arcsec)-spacing grid. 
The spectral intensities, expressed in the $T_{\rm MB}$ scale, are consistent 
 between our \NtHp (1--0) observations and theirs after adopting the historical main beam efficiency ($\eta_{\rm MB}$) of 0.78 \citep{Greve98} for their observations and the current $\eta_{\rm MB}$ of 0.85 for our new observations.
We also produce the \NtHp (1--0) integrated intensity map with CLASS (see Fig.\,\ref{fig:maps_L1498}i). 

We performed the \HtDp (\oHtDpgrd) and \NtHp (4--3) observations in December 2015
using the 16 pixels HARP receiver equipped on  JCMT \citep{Buckle09},
of which two pixels are non-functioning,
in frequency-switching mode. 
The HARP array was rotated by 45$^\circ$ during the observation.
In the JCMT-pointing grid shown in Fig.\,\ref{fig:pointing_L1498}, the spacing is 15\arcsec along the northeast--southwest direction and 30\arcsec along the northwest--southeast direction.
Data are converted to the CLASS format to be reduced with CLASS (folding and baseline subtraction). 

The \NtDp (1--0) and \DCOp (1--0) observations were carried out
in November 2014
using GBT. 
We used the MM1 and MM2 W-band dual 
polarization sub-band receivers in in-band frequency-switching mode with the Versatile GBT Astronomical Spectrometer (VEGAS) backend. 
The spectra were observed on the same northeast--southwest cut as the IRAM 30-m observations.
Data were 
preprocessed in the GBTIDL data reduction program and converted to the CLASS format for subsequent 
reduction. 
The data were reduced by applying a technique where no OFF observation is subtracted (total power mode) to gain $\sqrt{2}$ in sensitivity.
This technique utilizes two displaced spectra (ON and OFF observations) obtained in the frequency-switching mode by realigning and averaging these spectra, instead of subtracting and folding them in the standard reduction procedure, as explained in \citet{Pagani20}.
The above spectral observations were conducted together with the previous L\,1512 observations \citep{Lin20}. We refer readers to the observational details described by \citet{Lin20}.

\subsubsection{Calibration}\label{sec:obs_spec_cali}

Our spectral data are presented in the $T_{\rm A}^*$ scale throughout this paper instead of the $T_{\rm MB}$ scale (see Fig.\,\ref{fig:maincut_spectra_L1498}). 
For extended sources like L\,1498, the $T_{\rm MB}$ scale can introduce an over-correction for low main-beam efficiencies ($\eta_{\rm MB}$) \citep{Bensch01}, particularly in the 1.3--0.8~mm range observed with the IRAM 30-m telescope and also in the 4~mm range observed with the GBT (see Table\,\ref{tab:obs_L1498}). 
In such cases, 
the error beams can pick up a substantial fraction of the signal of extended sources.
Therefore, we directly analyze our data in the $T_{\rm A}^*$ scale in Sect.\,\ref{sec:analysis_L1498_RT} 
by considering the telescope beam response coupling to L\,1498 based on the main beam and error beam efficiencies of the IRAM-30m telescope (see the online table\footnote{https://publicwiki.iram.es/Iram30mEfficiencies}) and GBT (see Appendix\,\ref{app:Ta}). 
For the archival IRAM-30m \NtHp (1--0) and \CetO (1--0) data obtained by \citet{Tafalla04} (see Fig.\,\ref{fig:n2hp10_spectra_L1498} and the bottom row in Fig.\,\ref{fig:c18o10_spectra_L1498}),
we adopted the historical main and error beam efficiencies measured by \citet{Greve98}.
On the other hand, 
we consider only the main beam response coupling for our JCMT \HtDp (\oHtDpgrd) and \NtHp (4--3) submm observations due to the unavailability of error beam measurements for JCMT, as noted by \citet{Buckle09}.
Despite this limitation, we find that the \HtDp emission area is less extended (see the \HtDp spectra in Fig.\,\ref{fig:maincut_spectra_L1498} and Fig.\,\ref{fig:h2dp_spectra_L1498}), whereas the \NtHp (4--3) line is not detected. Hence, there is a less pronounced contribution from extended structures that the error beams would pick up.

\subsection{Continuum observations}

\begin{figure*}[t]
	\centering
	\includegraphics[width=\textwidth]{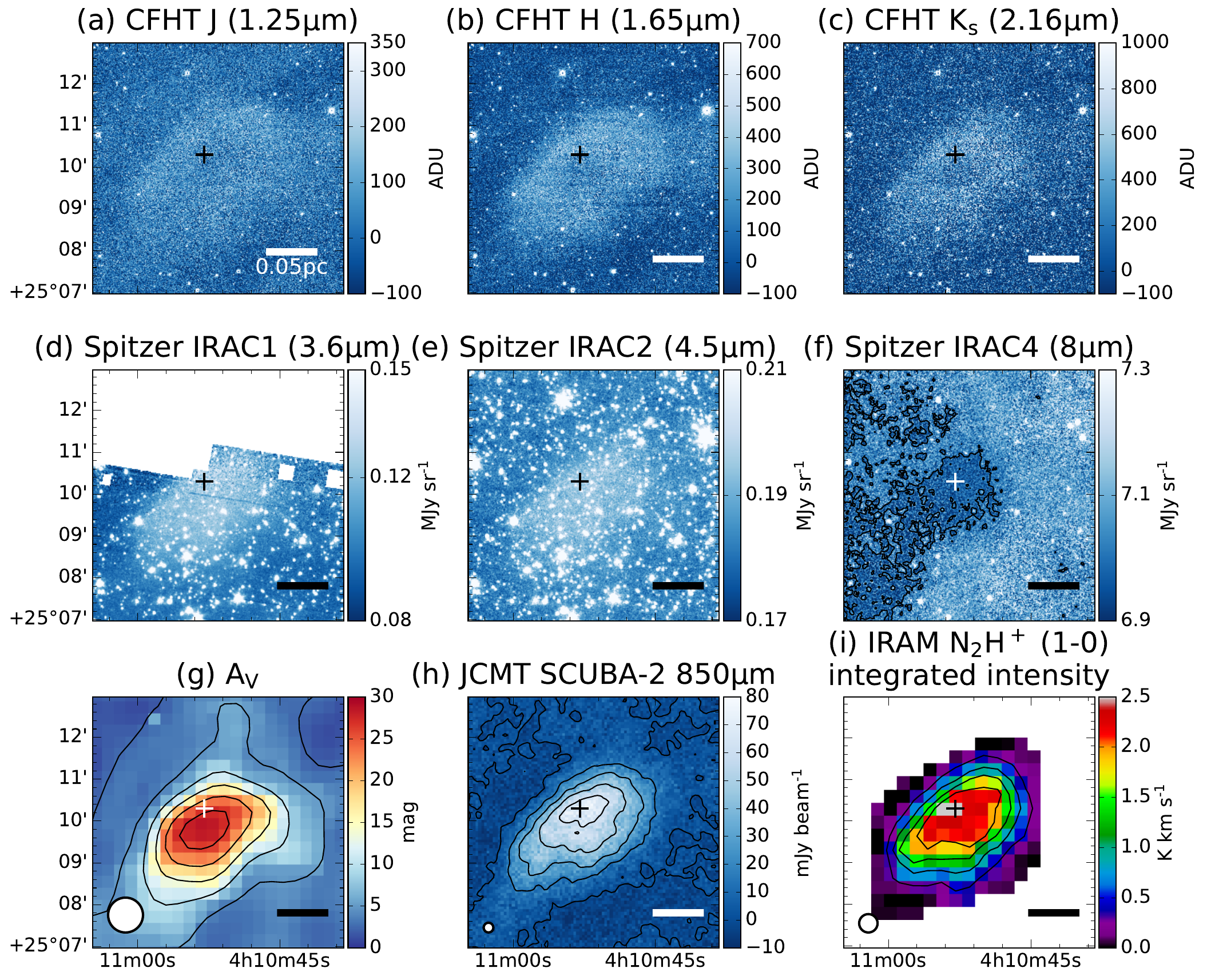}
	\caption{
    L\,1498 maps of continuum, extinction, and line integrated intensity.
    The CFHT NIR maps at (a) \J band, (b) \H band, and (c) \Ks band.
    Spitzer MIR maps at (d) IRAC1 band, (e) IRAC2 band, and (f) IRAC4 band with its contours at 7.0~MJy~sr$^{-1}$.
    (g) Visual extinction map with a beam size of 50\arcsec with contours at 2.5, 5, 10, 15, 20, 25~mag.
    h) JCMT SCUBA-2 850~\micron map with a beam size of 14\arcsec and the contours shown in Fig.\,\ref{fig:pointing_L1498}.
    (i) Integrated intensity maps of \NtHp $J$=1--0 from \citet{Tafalla04}, calculated within $V_{\rm LSR}$=[$-$0.5 km s$^{-1}$, 15.2 km s$^{-1}$]
    with its contours at 20, 40, 60, 80\% of its peak at 2.5 K km s$^{-1}$ and a beam size of 26\arcsec.
    The central cross in each panel indicates the center of L\,1498.
    The scale bars of 0.05 pc and \Av/millimeter-wavelength beam sizes 
    are denoted in the bottom right and bottom left corners, respectively.}
    \label{fig:maps_L1498}
\end{figure*}

\subsubsection{JCMT observations}

The Submillimeter Common-User Bolometer Array 2 \citep[SCUBA-2;][]{Holland13} 850~\micron observations of L\,1498 were performed in August 2021 and August 2022.
Four observations were taken in Band~2 weather (${0.05<\tau_{\rm 225GHz}<0.08}$)
under project codes M21BP043 and M22BP041 (PI: Sheng-Jun Lin),
as supplementary SCUBA-2 projects of the ongoing JCMT BISTRO-3 (B-fields In STar-forming Region Observations 3) survey (PI: Derek Ward-Thompson). 
The analysis of the SCUBA-2 photometric data combined with the polarimetric data from the BISTRO-3 survey will be presented in a forthcoming paper as part of the BISTRO-3 series.

Each SCUBA-2 observation 
consists of 44 minutes of integration
with the PONG-900 scan pattern, which fully samples a 15\arcmin diameter circular region.
The telescope beam size is 14\arcsec at 850~\micron.
To account for the typically faint and extended nature of starless cores, we processed the raw data using the \textit{skyloop} routine, employing a configuration file optimized specifically for extended emissions\footnote{https://www.eaobservatory.org/jcmt/2019/04/a-new-dimmconfig-that-uses-pca/}, provided by the
\textsc{SMURF} package in the \textit{Starlink} software suite \citep{Chapin13}.
Figures~\ref{fig:pointing_L1498} and \ref{fig:maps_L1498}h show the reduced data.
The map was gridded to 4\arcsec pixels and calibrated using a flux conversion factor (FCF) of 495~Jy~pW$^{-1}$ \citep{Mairs21}. The average rms noise in the central 15\arcmin field of the map on 4\arcsec pixels is found to be $\sim$4~\mJybm.

\subsubsection{Canada-France-Hawaii Telescope (CFHT) and Spitzer observations}

The CFHT Wide InfraRed CAMera (WIRCAM) was used with the wide filters \J, \H, and \Ks to observe the source on the night of 26 December 2013. 
Seeing conditions were typically 0\farcs8, and
the on-source integration times were typically 0.5 to 1 hour per filter, to reach a completion magnitude of 21.5 (\J band) to 20 (\Ks band).
We refer readers to the reduction details described by \citet{Lin20}.
Spitzer observations are collected from the Spitzer Heritage Archive
(SHA)\footnote{https://sha.ipac.caltech.edu/applications/Spitzer/SHA/}. 
L\,1498 was observed with Spitzer InfraRed Array Camera (IRAC)
in two programs: Program id. 94 (PI: Charles Lawrence) and Program id. 90109 (PI: Roberta Paladini); the second program occurred during the
warm period of the mission (only 3.6 and 4.5~$\mu$m channels still working). The data taken from Program id.\,94 were already discussed in
\citet{Stutz09}, while the data taken from Program id.\,90109 are analyzed in \citet{Steinacker15}. These two papers give the
observational details. During the warm mission, deep observations of the cloud were performed and a completion magnitude
of 18.5 in band IRAC2 (4.5~\micron) was reached.

\begin{figure*}[!t]
\vspace{-0.4cm}
	\centering
	\includegraphics[width=0.95\textwidth]{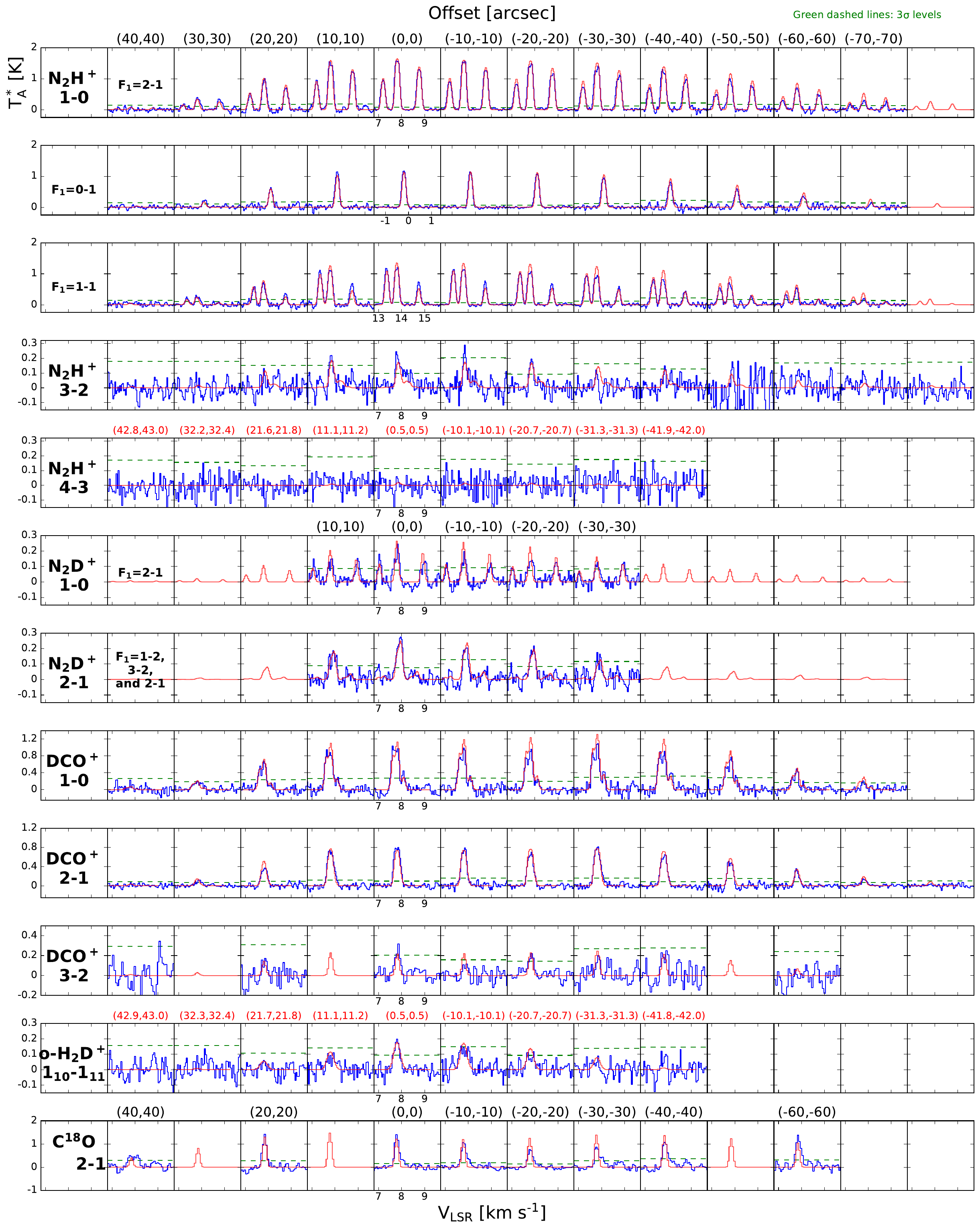}
	\caption{
    Spectral observations along the main cut compared to our best-fit radiative transfer model.
    The blue spectra show the observational data and the red spectra show the models. Each column corresponds to different offsets from the center of L\,1498 according to Fig.\,\ref{fig:pointing_L1498}. Each row shows a spectral line, except that the \NtHp (1--0) line is split into three rows corresponding to its different $F_1$-transition groups.
    The green dashed lines indicate the 3$\sigma$ noise level.
    Observational parameters are summarized in Table\,\ref{tab:obs_L1498}.}
    \label{fig:maincut_spectra_L1498}
\end{figure*}

\section{Results}\label{sec:results_L1498}

\subsection{Continuum maps}

\begin{figure*}[t]
	\centering
    \includegraphics[scale=0.3]{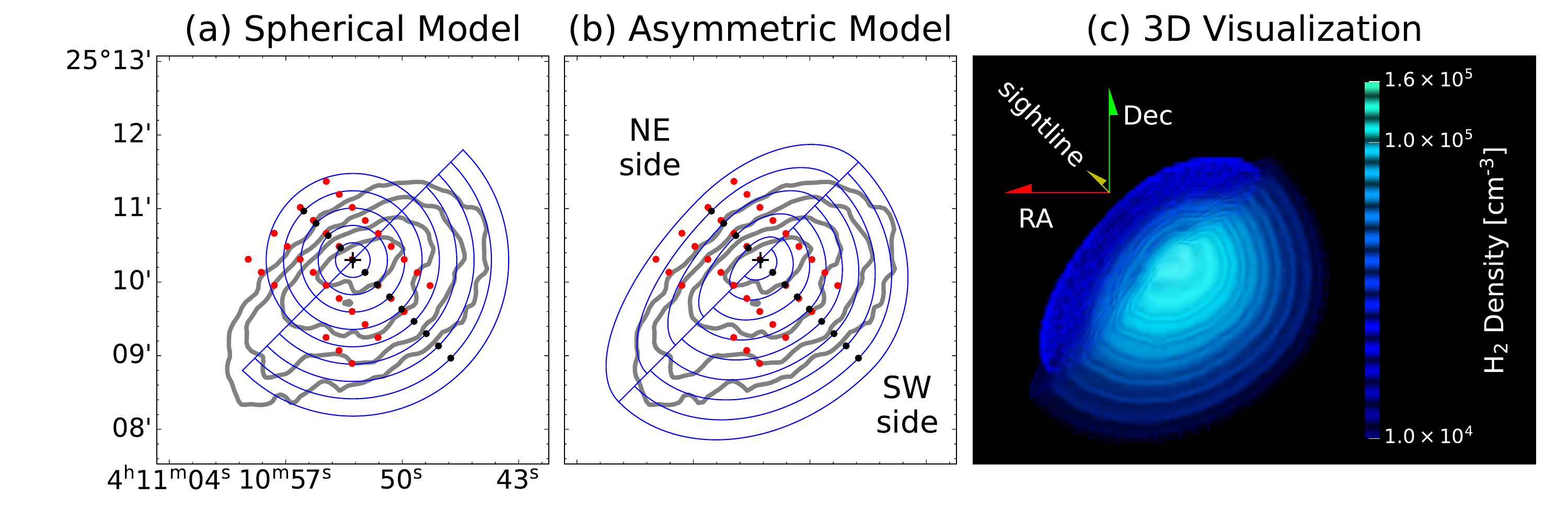}
    \caption{
    Onion-like physical models.
    (a) The spherical model with the layer width as $10\sqrt{2}$ arcsec ($\approx14.1$\arcsec and 1980~au at the distance of 140~pc, the same as the spacing of IRAM 30-m/GBT pointing observations) and (b) the asymmetric model overlaid on the contours of the 850~\micron map. The contour levels are 20\%, 40\%, 60\%, and 80\% of the peak at 78~\mJybm.
    (c) 3D visualization of the \Ht density distribution in the asymmetric model.}
    \label{fig:onion_model_L1498}
\end{figure*}

The continuum maps of L\,1498 at near-infrared (NIR), mid-infrared (MIR), and sub-millimeter (submm) wavelengths are shown in Fig.\,\ref{fig:maps_L1498}.
With the benefit of the deep NIR and MIR observations,
the cloudshine phenomenon \citep{Foster06} is detected at \J, \H, and \Ks bands, while 
the coreshine phenomenon \citep{Pagani10a, Steinacker10, Lefevre14} is detected at IRAC1 and IRAC2 bands.
This cloud/coreshine detection indicates the presence of dust grain growth \citep{Steinacker15}.
We analyze the dust extinction in the above deep \J, \H, \Ks, and IRAC2 images
to derive the visual extinction map with a beam size of 50\arcsec shown in Fig.\,\ref{fig:maps_L1498}g.
We attempt to derive the total column density all over the cloud using the \Av map with the benefit that the dust extinction is 
dependent on the dust density but independent of the dust temperature
\citep{Pagani04, Pagani15, Lefevre16}.
However, because of the high extinction toward the core center,
the lack of sufficient \Ks-band stars prevents us from deriving the actual \Av peak value but a lower limit of \Av at 25~mag toward the core center.
This \Av map is then provided as a constraint on the outer density profile derived by the \NtHp line emission (see Sect.\,\ref{sec:analysis_Av_L1498}).

The 850~\micron dust continuum emission observed by JCMT (Fig.\,\ref{fig:maps_L1498}h) is optically thin and sensitive to the cold dust in the core.
With a much smaller 850~\micron beam size of 14\arcsec compared to that of the \Av map, it clearly reveals the northwest--southeast-elongated core shape with a concave edge at the southern side.
Compared with the \NtHp (1--0) integrated line intensity map (Fig.\,\ref{fig:maps_L1498}i) with a beam size of 26\arcsec obtained from \citet{Tafalla04}, we can see that both the 850~\micron continuum emission and the \NtHp (1--0) emission peak at the same position and have similar emission distributions, which implies that both trace the same region.
Therefore, we use the 850~\micron continuum map and the \NtHp (1--0) integrated intensity map to
determine the core center of L\,1498 to be 
(RA, Dec)$_{\rm J2000}$ = (4$^{h}$10$^{m}$52\fs97, +25\degr10\arcmin18\farcs0).
We can see that the chosen core center also coincides with the absorption center on the IRAC4 image (Fig.\,\ref{fig:maps_L1498}f), where the absorption feature is associated with the inner region of the core \citep{Lefevre16}.
We denote the core center as a cross shown in each panel of Fig.\,\ref{fig:maps_L1498}.

\subsection{Molecular emission lines}\label{sec:results_mol_L1498}

Figure\,\ref{fig:pointing_L1498} shows the pointings of our observations.
Our IRAM 30-m and GBT pointing observations are performed along a northeast--southwest cut across the core center,
and one row in the JCMT pointings is overlaid on the same northeast--southwest cut.
We used this cut (hereafter, the main cut) to analyze the physical and chemical properties of L\,1498 (Sect.\,\ref{sec:analysis_L1498}).
Figure\,\ref{fig:maincut_spectra_L1498} shows the spectra of each emission line in the $T_{\rm A}^*$ scale along the main cut.

The L\,1498 core is quiescent. Each hyperfine component of \NtHp (1--0) and \NtDp (1--0) is well separated. 
Along the main cut, the coverage of the \NtHp (1--0), \NtHp (3--2), \DCOp (1--0), and \DCOp (2--1) spectra span across the whole core.
Including the \NtDp and o-\HtDp observations, we can see that the spectral intensities of these four cations peak toward the central region,
suggesting that these cations are tracers for the core region.
Meanwhile, \NtHp (4--3) shows no detection beyond the 3$\sigma$ significance because of the low temperature inside the core and low central density compared to the critical density for this line ($\sim2\times10^7$~cm$^{-3}$ at 10~K).
On the other hand,
the o-\HtDp (\oHtDpgrd) line is detected in a smaller elongated region with an extent of $\sim$30\arcsec along the main cut and $\sim$60\arcsec perpendicular to the main cut (also see Fig.\,\ref{fig:h2dp_spectra_L1498}).
This indicates that o-\HtDp traces the innermost region of the L\,1498 core.
We note that this is the first o-\HtDp (\oHtDpgrd) detection toward L\,1498.
\citet{Caselli08} conducted a survey of o-\HtDp (\oHtDpgrd) in nearby dense cores with the CSO 10.4-m antenna. However, L\,1498 is one of the two undetected targets out of ten starless cores in their sample, indicating that the deuterium fractionation in L\,1498 is relatively low compared with the other starless cores of their study.
Their non-detection is probably because 
their rms noise of $\sim$130~mK in the $T_{\rm MB}$ scale is larger than our rms noise of $\sim$50--85~mK in the same scale (see Table \,\ref{tab:obs_L1498}).
In addition, 
we also have a few \CetO (2--1) spectra observed along the main cut,
of which the intensity shows a clear decrease in the central region
although the minimum spectral intensity does not occur at the core center but at ($-$20\arcsec, $-$20\arcsec).

\section{Analysis}\label{sec:analysis_L1498}

\subsection{Objective and modeling strategy}

We aim to provide a 3D physical description of L\,1498 and 
estimate the chemical timescale of the deuteration fractionation.
To accomplish this goal, we utilized non-LTE radiative transfer modeling to analyze molecular line emissions, enabling us to evaluate the physical structure and molecular abundance profiles of the core. 
This abundance analysis is essential for estimating the chemical timescales of observed $X$(\NtDp)/$X$(\NtHp) ratios and other deuterated molecular abundances through time-dependent chemical modeling.

In the non-LTE radiative transfer modeling, we approximated the physical structure of L\,1498 as two half onion-like models stuck together, comprised of multiple concentric homogeneous layers.
The parameters in each layer are
number density ($n_{\rm H_2}$), 
gas kinetic temperature ($T_{\rm kin}$), 
relative abundances with respect to \Ht of the observed species ($X_{\rm species}=n_{\rm species}/n_{\rm H_2}$), 
turbulent velocity ($V_{\rm turb}$\footnote{$V_{\rm turb}$ is defined as the 3D isotropic non-thermal velocity dispersion. If $T_{\rm kin}$ and $V_{\rm turb}$ are constant along the line of sight, the FWHM of the line profile is given by $\delta v_{\rm FWHM}^2=8\log(2)\cdot\left(\frac{k_{\rm B}T_{\rm kin}}{m}+\frac{1}{3}V_{\rm turb}^2\right)$, where $k_{\rm B}$ is the Boltzmann constant and $m$ is the mass of the observed species.}), 
rotational velocity field ($V_{\rm rot}$), and 
radial velocity field ($V_{\rm rad}$).
The layer width is chosen to be $10\sqrt{2}$~arcsec ($\approx14.1$\arcsec and 1980~au at the distance of 140~pc), the same as the diagonal spacing of our IRAM 30-m/GBT pointing observations along the main cut. 
We therefore can determine these parameters at each layer sequentially from the outermost to the innermost layer by sampling sightlines of progressively decreasing radius along the main cut across L\,1498.

Despite observations indicating an infall motion within the L\,1498 envelope \citep{Lee01, Lee11, Magalhaes18}, the L\,1498 core remains quiescent, as evidenced by the single-peaked spectra observed in our study. Furthermore, the velocity gradient of \NtHp within this region is characterized by small magnitudes and seemingly random orientations, with a reported mean magnitude of 0.5$\pm$0.1~km~s$^{-1}$~pc$^{-1}$ and a direction of $9^\circ$$\pm$10$^\circ$ \citep{Caselli02}. Based on these observations, we have proceeded with the assumption that $V_{\rm rad}=0$ and $V_{\rm rot}=0$.

Then our procedure to determine these parameters at each layer is as follows. 
First, we used multi-transition \NtHp spectra to determine the $n_{\rm H_2}$, $T_{\rm kin}$, and $X$(\NtHp) profiles, assuming a uniform $V_{\rm turb}$ value. 
Second, we determined the abundance profiles of \NtDp, \DCOp, and o-\HtDp
by assuming that they share the same density, temperature, and kinematic properties of \NtHp. The details of the radiative transfer modeling are explained in Sect.\,\ref{sec:analysis_L1498_RT}.
With the obtained abundance profiles of \NtHp, \NtDp, \DCOp, and o-\HtDp, we employed a pseudo-time-dependent chemical model \citep{Pagani09b} to estimate the chemical timescales of each layer with our derived density and temperature profiles. 
In this adopted chemical model, the free parameters include the abundances of CO and \Nt, the initial OPR of \Ht (OPR$_{\rm intial}$(\Ht)), the average grain radius ($a_{\rm gr}$), and the cosmic ray ionization rate ($\zeta$). These parameters are either determined or assumed with chosen values during the modeling process. Further details of the chemical modeling are explained in Sect.\,\ref{sec:analysis_chem_L1498}.

We basically followed the procedure in \citet{Lin20}, with the main difference being the construction of an asymmetric onion-like model for L\,1498. 
Additionally, while \citet{Lin20} independently determined $n_{\rm H_2}$ using the NIR/MIR dust extinction measurements, we simultaneously determined $n_{\rm H_2}$ along with $T_{\rm kin}$, $X$(\NtHp), and $V_{\rm turb}$ by reproducing our multi-transition \NtHp spectra through radiative transfer modeling. 
This was due to the lack of sufficient \Ks-band stars, which prevented us from deriving the actual \Av peak value but allowed us to establish a lower limit of \Av toward the core center (see Sect.\,\ref{sec:analysis_Av_L1498}). Thus our dust extinction map served primarily as a constraint on the $n_{\rm H_2}$ profile at outer radii, where \Av can be determined. We will further discuss the agreement of the dust extinction with our $n_{\rm H_2}$ in Sect.\,\ref{sec:discussion_den_L1498}.

\begin{figure*}[!ht]
	\centering
	\includegraphics[scale=0.4]{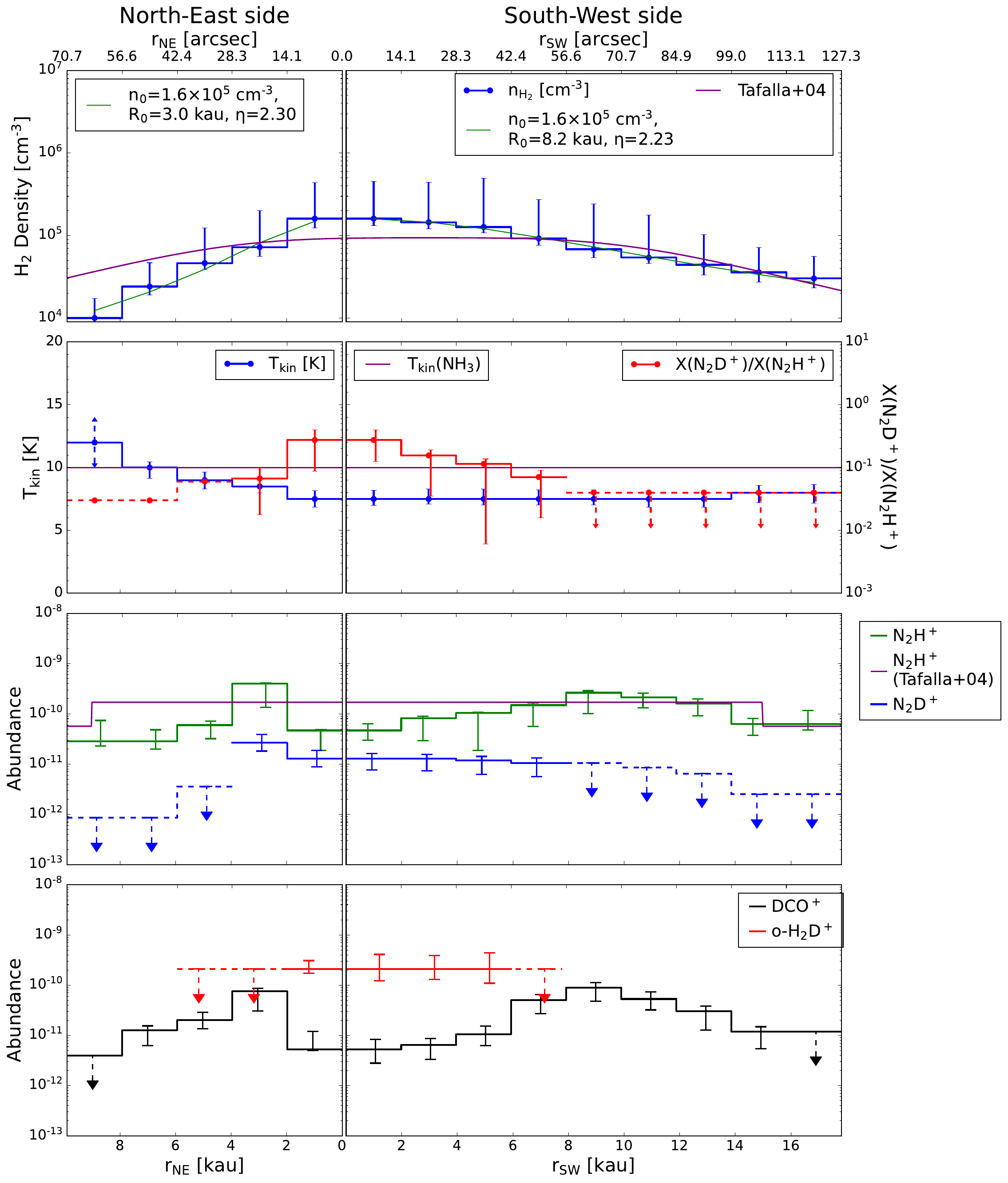}
	\caption{
    Physical and abundance profiles along the main cut.
    The profiles of number density ($n_{\rm H_2}$), kinetic temperature ($T_{\rm kin}$), the abundance ratio of \NtDp/\NtHp, abundances ($X_{\rm species}$) are the best-fit results from the non-LTE radiative transfer calculations.
    The Plummer-like profiles fitted with our density profiles are plotted as green curves and the corresponding Plummer parameters are annotated. 
    The purple solid curves show the $n_{\rm H_2}$, $T_{\rm kin}$(\NHth), $X$(\NtHp) profiles derived by \citet{Tafalla04}.}
    \label{fig:rad_model_L1498}
\end{figure*}

\begin{figure*}
	\centering
	\includegraphics[scale=0.34]{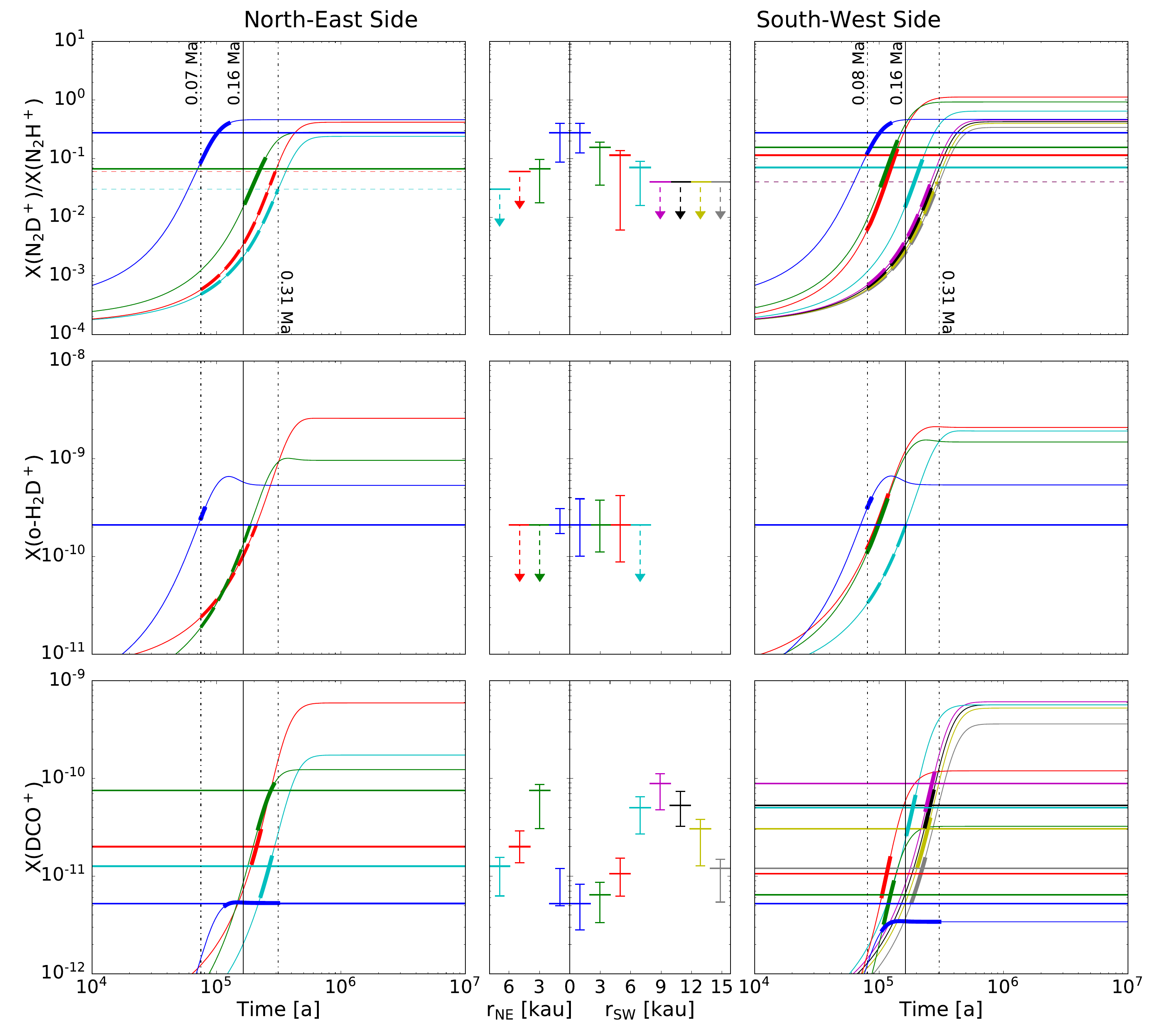}
	\caption{
    Chemical modeling of the abundance ratio of \NtDp/\NtHp and the abundances of
    o-\HtDp and \DCOp for each layer.
    \textit{Left and right columns}: Chemical model solutions (curves) and the observationally derived values (horizontal lines).
    \textit{Middle column}: Observationally derived profiles with uncertainties from Fig.\,\ref{fig:rad_model_L1498}.
    The model solutions and observed values are color-coded by different layers.  
    The models are
    calculated with an initial OPR(\Ht) of 3, a cosmic ray ionization rate ($\zeta$) of $3\times10^{-17}$\,s$^{-1}$, and
    an average grain radius ($a_{\rm gr}$) profile shown in Fig.\,\ref{fig:chem_model_L1498_CO_N2}.
    The two dashed-dotted vertical lines in each panel indicate a time range for which the model values cross the observations within their error bars, and we make such model curves thicker. 
    The thick model curve is displayed as dashed if the observation only has an upper limit.
    The solid vertical line indicates the lower limit on the core age of L\,1498 as 0.16~Ma.
    }
    \label{fig:chem_model_L1498}
\end{figure*}

\subsection{Radiative transfer applied to the onion-like physical model}\label{sec:analysis_L1498_RT}

Figure\,\ref{fig:onion_model_L1498}a and \ref{fig:onion_model_L1498}b show two onion-like models of L\,1498. One is 1D spherically dissymmetric and the other is 3D asymmetric (see Appendix\,\ref{app:asym_model}).
With our spectral observations along the main cut, we assumed that both models comprise nine and five layers
on the southwest and northeast sides, respectively. 
In order to reproduce our observed spectra,
we adopted a 1D spherically symmetric non-LTE radiative transfer code \citep[\texttt{MC};][]{Bernes79, Pagani07} for the 1D spherical case,
and the Simulation Package for Astronomical Radiative Transfer/Xfer code (\texttt{SPARX}\footnote{https://sparx.tiara.sinica.edu.tw/})
for the 3D asymmetric case.
In both codes, we turned on the hyperfine-line-overlapping feature, and provided
the updated hyperfine-line-resolved collisional rate coefficients \citep{Lique15, Pagani12, Lin20}.
We then obtained our modeled spectra calibrated in the $T_{\rm A}^*$ scale,
whereas the $T_{\rm MB}$ scale can introduce an overcorrection for low main-beam efficiencies, $\eta_{\rm MB}$ \citep{Bensch01}. Particularly in the 1.3--0.8~mm range observed with the IRAM 30-m telescope and also in the 4~mm range observed with the GBT, the error beams can pick up a substantial fraction of the signal of extended sources (see Sect.\,\ref{sec:obs_spec_cali}). This leads to overestimates of the temperature and/or density and also has an impact on the molecular abundance and the abundance ratio.
Calibrating in the $T_{\rm A}^*$ scale by considering the telescope coupling contributions from the main beam and error beams allows for a better representation of extended sources without having to handle the corrections of the error beam pick-up in the $T_{\rm MB}$ scale (see Appendix\,\ref{app:Ta}).

We search for the best-fit profiles of $n_{\rm H_2}$, $T_{\rm kin}$, abundances of our observed species, and $V_{\rm turb}$ with the 1D spherical model by our spectral fitting procedure. 
Initially, we ran \texttt{MC} 
to reproduce our \NtHp (1--0, 3--2, and 4--3) spectra across the main cut.
This allowed us to determine $n_{\rm H_2}$, $T_{\rm kin}$, $X$(\NtHp), and $V_{\rm turb}$ layer by layer, sequentially from the outermost to the innermost layer, for both the northeast and southwest sides.  
We found that a uniform $V_{\rm turb}$ of $0.09$~km~s$^{-1}$ can reproduce the spectral line widths along the main cut, with all spectra consistent with a systematic velocity of 7.8~km~s$^{-1}$.
These initial $n_{\rm H_2}$, $T_{\rm kin}$, and $X$(\NtHp) profiles 
served as the initial guess for refinement. 
Subsequently, we used the simulated annealing (SA) algorithm provided by the Modeling and Analysis Generic Interface for eXternal numerical codes \citep[\texttt{MAGIX};][]{Moller13} to optimize the fit of the \NtHp spectra, obtaining the best-fit profiles of $n_{\rm H_2}$, $T_{\rm kin}$, and $X$(\NtHp) from our initial profiles.
The 1$\sigma$ uncertainty of each parameter at every layer was determined using the interval nested sampling error estimation method provided by \texttt{MAGIX}, with the other parameters fixed. 
Afterward, we determined the abundance profiles of \NtDp, \DCOp, and o-\HtDp, along with their respective 1$\sigma$ uncertainties, by fitting the spectra of \NtDp (1--0, and 2--1), \DCOp (1--0, 2--1, and 3--2), and o-\HtDp (\oHtDpgrd).
This fitting process assumed identical $n_{\rm H_2}$ and $T_{\rm kin}$ profiles, as well as a constant $V_{\rm turb}$, as for \NtHp.
Because of our shorter spatial coverage in the \NtDp observation, we limit $X$(\NtDp) in the outer layers with $X$(\NtDp)/$X$(\NtHp) lower than the observed minimum $X$(\NtDp)/$X$(\NtHp) of 0.07.
The best-fit central $n_{\rm H_2}$ was found to be $1.6_{-0.3}^{+3.0}\times10^5$~cm$^{-3}$, and the central $T_{\rm kin}$ is 7.5$_{-0.5}^{+0.7}$~K.

\begin{figure*}
	\sidecaption
	\includegraphics[width=12cm]{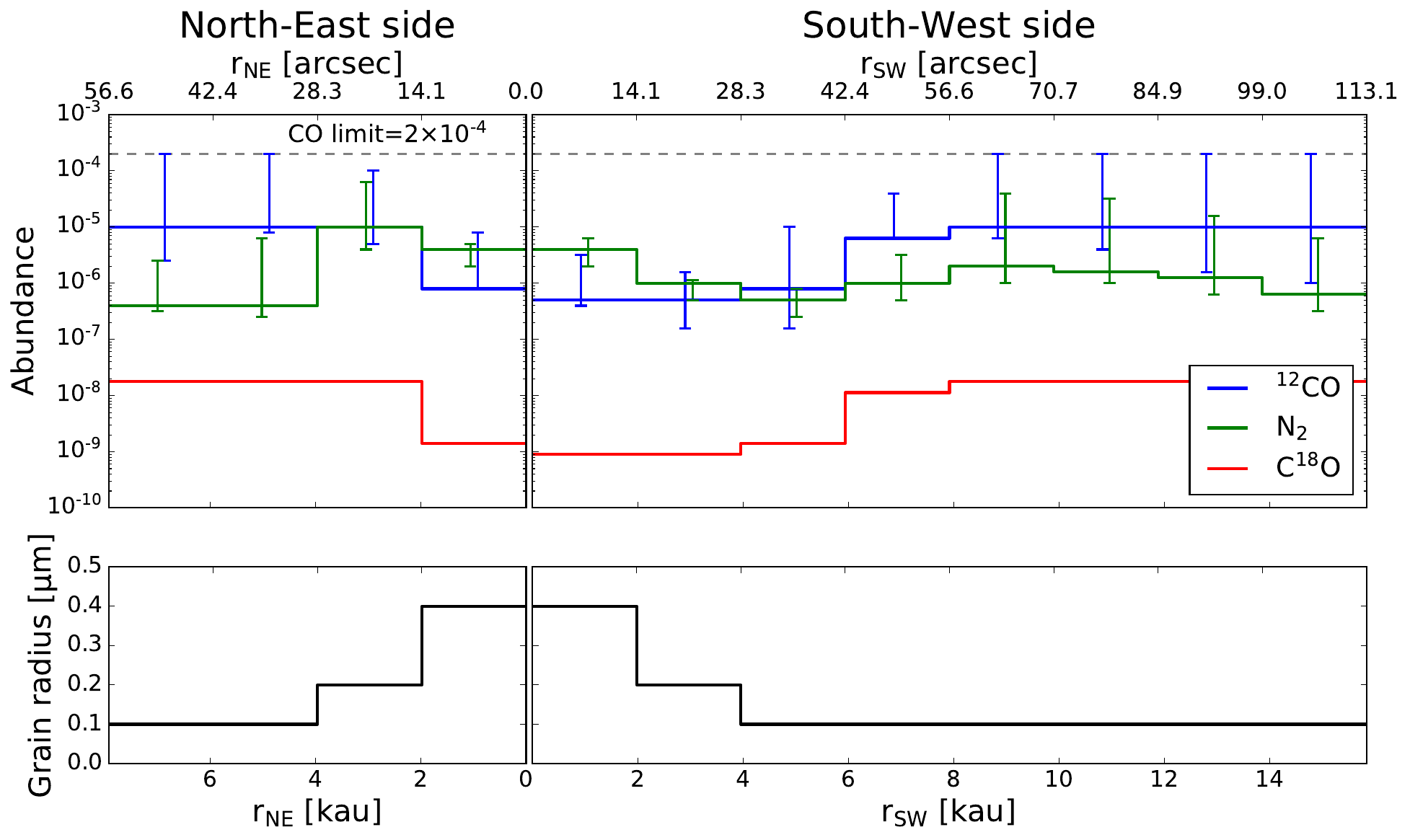}
	\caption{
    Profiles of the abundances CO, N$_2$, and the grain radius ($a_{\rm gr}$).
    The $X$(\twvCO), $X$(N$_2$), and $a_{\rm gr}$ profiles are the best-fit results from the chemical modeling. 
    The $X$(\CetO) profile is obtained by assuming a \twvCO/\CetO 
    abundance ratio of 560 or a constant value of $1.9\times 10^{-8}$.}
    \label{fig:chem_model_L1498_CO_N2}
\end{figure*}

Next, we transform the above best-fit profiles from the 1D spherical model to the 3D asymmetric model
because carrying out a non-LTE calculation with a 3D asymmetric model repeatedly in the fitting is much more numerically expensive than the 1D case.
Our 3D asymmetric model was designed to follow the emission morphologies of the 850~\micron continuum and the \NtHp (1--0) integrated intensity.
We set the southwest sides in both models to have identical physical parameters and abundance profiles. 
For the northeast side, the spherical model was stretched along the sightline direction by a factor of $\sim$2 to match the thickness of the southwest side along the line of sight (see Appendix\,\ref{app:asym_model}). Keeping the same northeast profiles in the asymmetric model results in too strong intensities in the spectra because the molecular abundances are doubled.
We aim to keep the continuity in the $n_{\rm H_2}$ and $T_{\rm kin}$ profiles across two sides, so we simply divided the abundances in the northeast sides by a common factor of $\sim$2 to make the modeled spectra fit with the observations.
We note that our intention is to obtain a simultaneous fit that is generally consistent with the observational spectra
since performing a complete search of the entire parameter space
for the best possible fit is not achievable.

Figure\,\ref{fig:rad_model_L1498} shows our best-fit profiles of $n_{\rm H_2}$, $T_{\rm kin}$, abundances of the above four cations, and 
the \NtHp deuteration ratio, $X$(\NtDp)/$X$(\NtHp), along the main cut across the center of our asymmetric onion-like model, while 
Figure\,\ref{fig:onion_model_L1498}c shows the 3D visualization of the $n_{\rm H_2}$ distribution.
We also fit our $n_{\rm H_2}$ profiles with
the Plummer-like profile with
$n_{\rm H_2}(r) = n_0/(1+(r/R_0)^\eta)$, where $n_0$, $R_0$, and $\eta$ are central density, flattening radius, and the power-law index, respectively.
The fitted Plummer-like profiles are shown with green curves and the fitted parameters are annotated 
($n_0 = 1.6\times 10^5$~cm$^{-3}$, $R_0=8.2$~kau and $\eta=2.23$ for the southwest profile, and $R_0=3.0$~kau and $\eta=2.30$ for the northeast profile)
in the top row in Fig.\,\ref{fig:rad_model_L1498}.
The above best-fit profiles are also numerically listed in Table\,\ref{tab:rad_model_L1498}.
In addition, in Fig.\,\ref{fig:rad_model_L1498}, the $n_{\rm H_2}$, $T_{\rm kin}$, and $X$(\NtHp) profiles of the spherical model found by \citet{Tafalla04} are also shown in purple. 
Since their model center was chosen at a different position with an offset of ($\Delta\mbox{RA}=-10$\arcsec, $\Delta\mbox{Dec}=-20$\arcsec) with respect to our model center, 
we compute these purple curves along our main cut (i.e., a secant line of their spherical model) using their parameters (see their Table 1 and 3).
We will further discuss the discrepancy between our and their models in Sect. \ref{sec:discussion_L1498}.
Finally, the best-fit modeled spectra of the four cations along the main cut are shown in red in
Fig.\,\ref{fig:maincut_spectra_L1498}. 
In addition, Figure \ref{fig:n2hp10_spectra_L1498} shows our best-fit modeled \NtHp (1--0) spectra overlaid on the data from \citet{Tafalla04}, while
Figure \ref{fig:h2dp_spectra_L1498} shows our best-fit modeled  o-\HtDp (\oHtDpgrd) spectra overlaid on our full o-\HtDp data.
Our reproduced spectra are in good agreement with the observations, suggesting that our model is globally valid for the L\,1498 core.

\begin{figure*}
	\centering
    \includegraphics[scale=0.43]{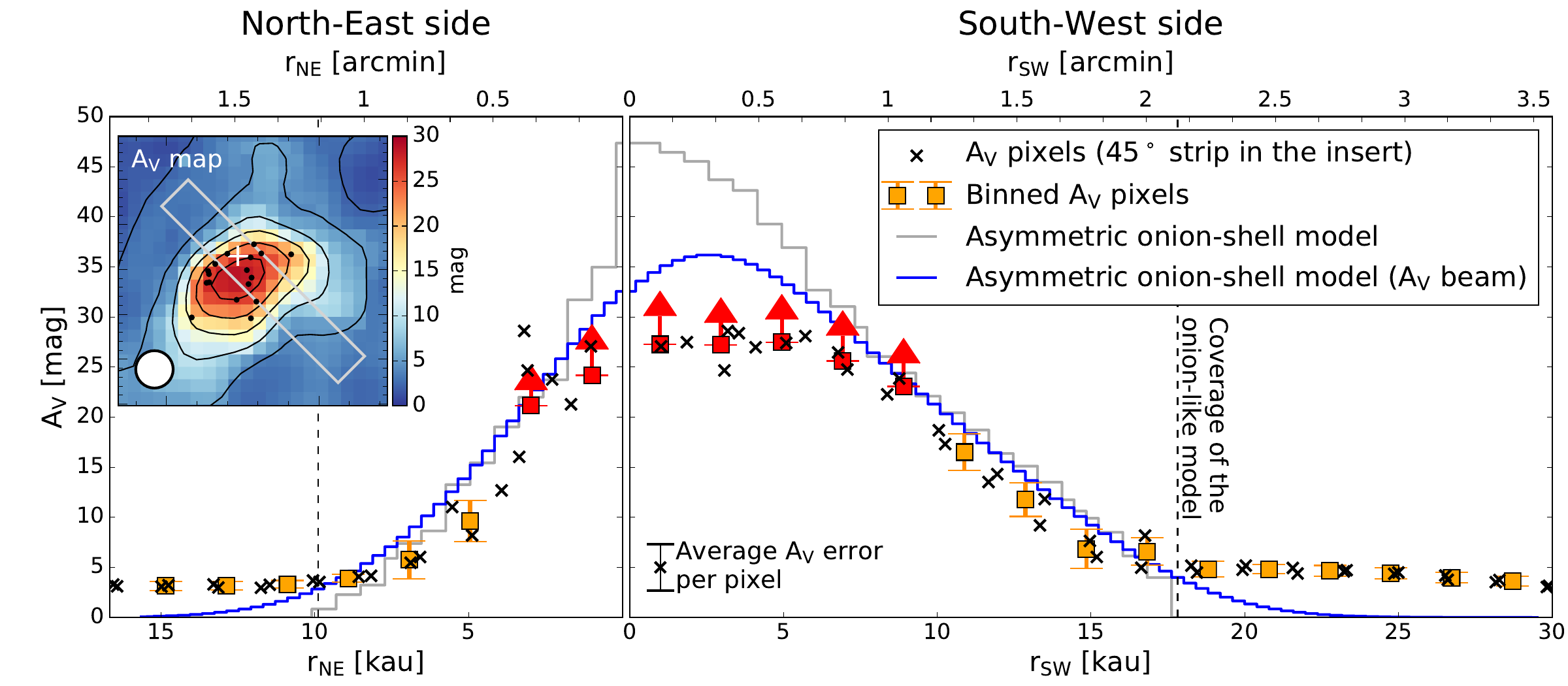}
    \caption{
    Comparison of the NICER-derived \Av profile and the \Av profile derived from the column density from the best-fit asymmetric onion-like model along the main cut.   
    The black crosses are the NICER-derived \Av values at each pixel in the 50\arcsec-wide strip on the \Av map shown as the insert.    
    The orange/red squares with error bars show the radially averaged \Av profile
    with the 14.1\arcsec-bin from the strip, where the central region is represented as lower limits due to the lack of the \Ks-band detections.
    The grey step curve shows the \Av profiles converted from our onion-like model, while the blue step curve is convolved with the \Av-beam of 50\arcsec.
    The coverage of our onion-like model is shown by black dashed lines.
    The insert shows the NICER-derived \Av map with the 50\arcsec-beam (Fig.\,\ref{fig:maps_L1498}g) overlaid with black dots, indicating the stars only detected in the IRAC1/2 bands but with the artificial \Ks-band detection (see Sect.\,\ref{sec:analysis_Av_L1498}).}
    \label{fig:Av_fitting_L1498}
\end{figure*}

\subsection{Time-dependent chemical model}\label{sec:analysis_chem_L1498}

We estimated the chemical timescale of each layer in the onion model of L\,1498 with our derived abundance, $n_{\rm H_2}$ and $T_{\rm kin}$ profiles.
Our method adopted a pseudo-time-dependent gas-phase deuterium chemical model \citep{Pagani09b} to simultaneously reproduce the deuteration ratio, $X$(\NtDp)/$X$(\NtHp), 
and the abundances of \NtHp, \NtDp, \DCOp, and o-\HtDp in each layer.
This deuterium chemical model is specialized for starless cores in that 
each spin-state of \Ht and \Hthp isotopologues are included and 
the complete depletion in heavy species, except for CO and \Nt, is assumed.
In our model, we did not include the nitrogen chemistry.
CO and \Nt would quasi-instantaneously reach chemical equilibrium with our observed species
because our chemical network is very small.
On one hand, CO and \Nt are the parent molecules reacting with the \Hthp isotopologues to form the \NtHp and HCO$^+$ isotopologues (i.e., the daughter cations); 
on the other hand, dissociative recombination of these daughter cations would convert them back to CO and \Nt.
The abundances of CO and \Nt are left as free parameters in each layer, so 
we could derive the current CO and \Nt abundance profiles.
The other free parameters are the initial OPR of \Ht (OPR$_{\rm intial}$(\Ht)), the average grain radius ($a_{\rm gr}$), and the cosmic ray ionization rate ($\zeta$).

Figure\,\ref{fig:chem_model_L1498} shows the best-fit chemical model solutions of $X$(\NtDp)/$X$(\NtHp), $X$(o-\HtDp), and $X$(\DCOp) for each layer at the northeast and southwest sides. 
The outermost layers were not modeled simply because their large observational error bars cannot well-constrain the chemical solutions.
For these chemical solutions, the initial OPR of \Ht is assumed to be its statistical ratio of 3.
We adopted the $\zeta$ of $3\times10^{-17}$ s$^{-1}$ throughout the chemical model, which is the best value found by \citet{Maret13} for their chemical model to reproduce the \CetO (1--0) and H$^{13}$CO$^+$ (1--0) integrated emission in L\,1498. 
With the detection of coreshine, one would expect that larger dust grains ($a_{\rm gr}>0.1$ \micron) appear in the core \citep{Pagani10a, Steinacker10, Lefevre14}. 
Although \citet{Steinacker15} cannot well constrain the maximum grain radius in L\,1498 with the coreshine modeling on the IRAC1/2 images, they suggested that the maximum grain radius seems to be greater than $\sim$0.3~\micron.
On the other hand, \citet{Maret13} found that in their chemical model, a power-law grain size distribution grown from the typical grain radius of 0.1~\micron to a maximum grain radius of 0.15~\micron is sufficient for reproducing the \CetO and H$^{13}$CO$^+$ line emissions, suggesting that the grain growth does not yet significantly occur everywhere in the entire core.
The apparent contradiction of the maximum grain radius reported by \citet{Maret13} and that reported by \citet{Steinacker15} arises because the low-$J$ transitions of \CetO and H$^{13}$CO$^+$ are optically thick, and as such, they do not trace the core center but rather the low-density outskirts. 
Thus, these indicate that $a_{\rm gr}\gtrsim0.3$~\micron at the center of L\,1498 while $a_{\rm gr}\approx0.1$~\micron at the outskirts.
The bottom row in Fig.\,\ref{fig:chem_model_L1498_CO_N2} shows the profiles of the grain radius in our chemical model.
In order to make the chemical solutions match the observational abundances,
we found that the grain radius should be at least 0.4~\micron in the innermost layer, and 0.2~\micron in the second layer. 
For the rest of the layers, we found matched chemical solutions with the typical grain radius of 0.1~\micron.
With the above conditions, we found that the observational abundances of all layers are matched with chemical solutions from 0.07 to 0.31~Ma on the northeast side, and from 0.08 to 0.31~Ma on the southwest side.

The top row in Fig.\,\ref{fig:chem_model_L1498_CO_N2} shows the CO and N$_2$ abundance profiles from our chemical modeling. Their best-fit values are listed in Table\,\ref{tab:rad_model_L1498}.
Error bars show the abundance ranges of the possible chemical solutions at each layer
but not every combination of $X$(CO) and $X$(\Nt) can make the chemical solutions fit the observational cation abundances.
We assume a constant $^{12}$C$^{16}$O/$^{12}$\CetO ratio of 560, which is comparable to the local $^{16}$O/$^{18}$O ISM ratio of 557$\pm$30 \citep{Wilson94}, to derive a \CetO abundance profile.
The \CetO line emission from the central depleted region is too weak to be constrained by radiative transfer calculation. Our chemical model serves as an appropriate approach.
At the outer layers, $X$(CO) is less constrained since deuteration is only an upper limit in the chemical models. By contrast, \CetO is not heavily depleted and its $J$=2--1 line emission remains thinner
and thus the radiative transfer modeling is applicable.
We then used our \CetO (2--1) observation to further determine the outer CO abundance (southwestern layer 4--8, and northeastern layer 2--4).
With radiative transfer calculations, we found an outer $X$(\CetO) of $1.9\times10^{-8}$ can roughly make the \CetO (2--1) spectrum models fit with observations (the last row in Fig.\,\ref{fig:maincut_spectra_L1498}), validating our chemical modeling results.
We then set the outer $X$(\twvCO) as $1.0\times10^{-5}$.

We noted that the \CetO (2--1) spectra around the pointing from ($-$10\arcsec, $-$10\arcsec) to ($-$40\arcsec, $-$40\arcsec) are somewhat lower in intensities compared to our  modeled \CetO spectra.
It seems to be caused by a more considerable depletion along the line of sight toward these positions.
It is possible that the \CetO depletion center shifted outward from the core center that we determined by the emission peak position shared by 850~\micron continuum, \NtHp and o-\HtDp data 
\citep[also see the \CetO $J$=1--0 and 2--1 integrated intensity maps shown in Fig.\,3 from ][]{Tafalla04}. 
This would hint that 
the CO isotopologues have a more complex spatial cloud structure along the line of sight.

\subsection{Visual extinction}\label{sec:analysis_Av_L1498}

Similar to the procedure in \citet{Lin20}, 
we used the NIR/MIR images (Fig.\,\ref{fig:maps_L1498}) to derive the visual extinction map with the NICER method \citep{Lombardi01}.  
We used the \J, \H, and \Ks images with the \Rv= 3.1 dust models from \citet{Weingartner01} to make an envelope extinction map tracing the L\,1498 envelope. 
On the other hand, we used the \H, \Ks, and IRAC2 images with the \Rv= 5.5B dust models from \citet{Weingartner01} to make a core extinction map tracing the L\,1498 core.
However, the lack of sufficient stars with the color excess $E_{K_s-{\rm IRAC2}}$ measurements prevents us from deriving the extinction at the core center.
Although many stars appear on the IRAC1/2 images, only a few stars are detected in the \Ks band in the core region,
with which we found that the magnitude differences of \Ks and IRAC2 bands ($m_{K_s}-m_{\rm IRAC2}$) for these stars approach 2.5~mag toward the center. This magnitude difference is already larger than the difference between the completion magnitudes of \Ks and IRAC2 bands (20 and 18.5~mag, respectively), which explains the disappearance of the \Ks-band counterparts.
Therefore, we assume that the \Ks-band magnitude of stars only detected in IRAC2 bands have $m_{K_s}\geq m_{\rm IRAC2} + 2.5$ mag, and we generated the artificial \Ks-band detection by  
\begin{enumerate}
    \item selecting stars that are (a) detected in both IRAC1 and IRAC2 bands to ensure it is not due to contamination in a single band and  
    (b) with the IRAC2 magnitude uncertainty less than 0.2 mag, and
    \item assigning their \Ks-band magnitude as $m_{\rm IRAC2} + 2.5$~mag, i.e., the brightest limit from the above assumption, resulting in a lower limit of $E_{K_s-{\rm IRAC2}}$ and \Av.
\end{enumerate}
After this process, the density of stars with color excess measurements across L\,1498 allowed us to convolve the reddening data with a 50\arcsec-Gaussian beam to produce the core and envelope extinction maps.
We combined the core map and envelope map by merging them at 5 mag and the final \Av map has been shown in Fig.\,\ref{fig:maps_L1498}g and the insert of Fig.\,\ref{fig:Av_fitting_L1498}. 
Stars only detected in the IRAC1/2 bands and with the artificial \Ks-band detection are displayed as black dots on the insert of Fig.\,\ref{fig:Av_fitting_L1498}. 
Figure \ref{fig:Av_fitting_L1498} also shows the radially averaged \Av profiles along the main cut within the 50\arcsec-wide strip.
As a result, the \Av values in the central region should be interpreted as the lower limit and therefore our result suggests that \Av $\gtrsim 25$ mag at the 50\arcsec-beam.

\section{Discussion}\label{sec:discussion_L1498}

As shown in Sect.\,\ref{sec:analysis_L1498}, we built an asymmetric onion-shell model of L\,1498 to evaluate its physical structure and chemical abundances  
shown in Figs.\,\ref{fig:rad_model_L1498} and \ref{fig:chem_model_L1498_CO_N2}. 
Our model can well reproduce the observed spectra along the main cut shown in Fig.\,\ref{fig:maincut_spectra_L1498} via the non-LTE radiative transfer calculations. 
In this section, we compare our findings with other studies and address the core age of L\,1498.

\subsection{Density and kinetic temperature}\label{sec:discussion_den_L1498}

With our non-LTE radiative transfer modeling on the \NtHp emission line spectra, we find 
a central $n_{\rm H_2}$ of $1.6_{-0.3}^{+3.0}\times10^{5}$~cm$^{-3}$ and $T_{\rm kin}$ of 7.5$_{-0.5}^{+0.7}$~K,
and a peak $N_{\rm H_2}$ of $4.5\times10^{22}$~cm$^{-2}$ toward the L\,1498 core.
With the submm data,
\citet{Tafalla04} derived $N_{\rm H_2}=\mbox{(3--4)}\times10^{22}$~cm$^{-2}$ with $T_{\rm d}=10$~K using the IRAM 30-m MAMBO 1.2~mm map.
In contrast, our $N_{\rm H_2}$ peak value is slightly larger than that of \citeauthor{Tafalla04}. 
This discrepancy could be due to our different approach using the radiative transfer modeling of the line emission instead of the continuum emission.
The crucial uncertainty in the non-LTE line modeling is the collisional coefficients, which is, however, much smaller than the uncertainty in the dust opacity at the submm wavelength.
On the other hand, \citeauthor{Tafalla04} adopted a uniform $T_{\rm kin}$ of 10~K derived from \NHth (1,1) and (2,2) lines as $T_{\rm d}$ by assuming the gas and dust are coupled. 
However, the lower critical density of these \NHth lines \citep[$\sim2\times10^3$~cm$^{-3}$;][]{Pagani07} compared to the \NtHp $J$=1--0 line ($1.3\times10^5$~cm$^{-3}$ at 10~K) would suggest that \NtHp can trace the inner $T_{\rm kin}$ better than \NHth.
Therefore, using 10~K may underestimate the density at the core center. 

Another distinction is that our onion model is asymmetric, while 
\citeauthor{Tafalla04}'s model is spherically symmetric.
The top row in Fig.\,\ref{fig:onion_model_L1498} shows the comparison between our number density profiles and those of \citeauthor{Tafalla04}. 
We can see that our $n_{\rm H_2}$ profiles and theirs converge toward the outer southwestern region.
While the spherical onion model adopted by
\citeauthor{Tafalla04} provides a good approximation for deriving an averaged density profile of the elongated L\,1498 core,
Their model center ($\Delta\mbox{RA}=-10$\arcsec, $\Delta\mbox{Dec}=-20$\arcsec with respect to ours) was based on the centroid of their 1.2~mm emission map rather than the emission peak.
In contrast, our model center is situated at the JCMT SCUBA-2 850~\micron continuum, \NtHp (1--0), and o-\HtDp (\oHtDpgrd) emission peaks, including their 1.2~mm continuum peak.
In this case, our chosen model center would allow us to capture the peak $N_{\rm H_2}$ rather than average it out with the surrounding region.

Meanwhile, our NIR and MIR dust extinction measurements provide another constraint on the density in L\,1498.
Different from the dust emission measurement, 
the dust extinction is independent of $T_{\rm d}$, and the NIR/MIR dust opacity is better determined than the submm dust opacity \citep{Pagani15, Lefevre16}.
Figure\,\ref{fig:Av_fitting_L1498} shows the comparison of the \Av profiles derived with the observational data and our onion model along the main cut.
The orange/red squares represent the extinction at the 50\arcsec-beam of the whole L\,1498 cloud (i.e., the core and envelope) 
derived from NIR/MIR images
and the central \Av value is estimated as the lower limit (\Av $\gtrsim 25$ mag; see Sect.\,\ref{sec:analysis_Av_L1498}).
The grey and blue step curves represent the \Av values contributed by the L\,1498 core, which
are converted from our asymmetric onion-like model by 
$N_{\rm H_2}/A_{\rm V}=9.4\times10^{20}$ cm$^{-2}$ mag$^{-1}$ \citep[\Rv= 3.1;][]{Bohlin78}, and the blue curve is convolved with the \Av beam size of 50\arcsec for the comparison with the data.
Our onion model represents the core region because it was constructed from the \NtHp data, a molecule confined in the core region.
We can see that the \Av profile of our onion model matches well with the data in the core region. 
The \Av profile of the data consists of the extinction from the L\,1498 core and from an envelope with \Av~$\approx$~3~mag.
Since our observations of \NtHp, \NtDp, \DCOp, and o-\HtDp are associated with the core as these molecules are mostly confined in the core region, we do not need to include the envelope to reproduce their observational spectra shown in Fig.\,\ref{fig:maincut_spectra_L1498}. In Appendix \ref{app:all_spectra_L1498_c18o}, we show that including an envelope with \Av = 3~mag in the radiative transfer model can still reproduce our \CetO (2--1) spectra (the bottom row in Fig.\,\ref{fig:maincut_spectra_L1498}), whereas it is necessary for reproducing the \CetO (1--0) spectra obtained by \citet{Tafalla04} with a lower critical density.

With the interferometer, the ALMA-ACA 1~mm continuum survey, FREJA \citep{Tokuda20}, found no substructure at the 1~kau scale in the central region of L\,1498, suggesting that the central density structure is very flat (characterized by a Plummer-like flattening diameter greater than 5~kau) and an upper limit on $n_{\rm H_2}$ of about $3\times10^5$ cm$^{-3}$. 
This non-detection is in agreement with the $n_{\rm H_2}$ profile evaluated from our onion model with
the central density of $1.6_{-0.3}^{+3.0}\times10^{5}$~cm$^{-3}$ and the central flattened region size of $\sim$11~kau ($R_{0, \rm{SW}}+R_{0, \rm{NE}}$; see the Plummer-like parameters shown on the top row in Fig.\,\ref{fig:onion_model_L1498}). 
Therefore, our solution from the non-LTE \NtHp radiative transfer modeling is compatible with the NIR/MIR dust extinction measurement as well as the 1mm continuum interferometry observation.

\subsection{Cation abundance profiles} \label{sec:discussion_ion_L1498}

We find that \NtHp is significantly depleted toward the L\,1498 center with a depletion factor of 5.6$_{-4.0}^{+2.1}$ with respect to the maximum $X$(\NtHp) at the fifth southwestern layer,
and another depletion factor of 8.5$_{-7.6}^{+0.5}$ with respect to the maximum at the second northeastern layer (see Fig.\,\ref{fig:rad_model_L1498}). Since the northeastern depletion feature is only resolved by two layers in the onion model, we take the depletion factor measured on the southwest side as the representative measurement in the following discussion (same for the other molecules).
Our finding of the \NtHp depletion
is opposite to the \NtHp enhancement by a factor of 3 reported by \citet{Tafalla04} with their radiative transfer modeling. 
Our central \NtHp abundance is $4.7\pm1.7\times 10^{-11}$, while their value is $1.7\times 10^{-10}$.
This discrepancy could be due to their different physical model and the approximation made in their radiative transfer calculations.
\citet{Tafalla04} derived a central $n_{\rm H_2}=9.4\times10^4$~cm$^{-3}$, a factor of 1.7 lower than our $1.6_{-0.3}^{+3.0}\times10^{5}$~cm$^{-3}$, and a constant $T_{\rm kin}$(\NHth)~=~10~K, higher than our 7.5$_{-0.5}^{+0.7}$~K 
(see their $n_{\rm H_2}$, $T_{\rm kin}$, and \NtHp abundance profiles on Fig.\,\ref{fig:onion_model_L1498}).
As previously mentioned, using the temperature derived from the \NHth lines could be biased by the warmer outer layers and thus one could overestimate the central temperature.
In addition, they omitted line overlap from their radiative transfer calculation in their customized version of the \texttt{MC} code, and the hyperfine-structure-resolved collisional coefficients were just not yet available.
As a result, the peak intensities of their best-fit \NtHp (1--0) spectra were too strong by $\sim$60\% compared to data, but even with our current \texttt{MC} code, adopting their physical model would still result in $\sim$30\% stronger intensities at the peaks.
In terms of the line-of-sight-integrated measurements, 
our averaged $X$(\NtHp) and \NtHp column density ($N_{\rm N_2H^+}$) toward the core center are $1.2\times10^{-10}$ and $5.2\times10^{12}$~cm$^{-2}$, respectively.
Our $N_{\rm N_2H^+}$ value is of the same order of magnitude compared with $1.7\pm0.7\times10^{12}$~cm$^{-2}$ reported by \citet{Crapsi05} despite their more simplistic LTE calculations.

Interestingly, both \NtHp depletion and enhancement are seen from chemo-dynamical modeling.
\citet{Aikawa05} performed a self-consistent calculation with a quasistatically contracting Bonner-Ebert sphere, and found that \NtHp is enhanced at a central $n_{\rm H_2}$ similar to L\,1498 ($1.5\times10^5$ cm$^{-3}$) and later becomes depleted at higher densities.
\citet{Holdship17} and \citet{Priestley18} 
used an analytical approach for the Bonner-Ebert density evolution, but found that the \NtHp depletion starts earlier in their calculation. \citet{Holdship17} found an \NtHp depletion factor of $\sim$50 when $n_{\rm H_2}$ reaches \citet{Tafalla04}'s best-fit value for L\,1498. 
Although this discrepancy might relate to their N chemistry details in the models, the depletion case would be preferred by our findings.

In starless cores, the depletion of the \NtHp and HCO$^+$ isotopologues is a result of the freeze-out of their parent molecules (\Nt and CO) in the core center.
In L\,1498, we find depletion factors of 5.6$_{-4.0}^{+2.1}$ for \NtHp and 17$\pm$11 for \DCOp. 
However, in contrast, \NtDp does not exhibit significant depletion toward the core center (see Fig.\,\ref{fig:rad_model_L1498}). 
Here, we do not interpret the inward decrease in $X$(\NtDp) on the northeastern side as significant depletion. 
This may be due to the limited coverage of our spectral observations in the northeastern area (see Fig.\,\ref{fig:maincut_spectra_L1498}). 
The lack of significant depletion in \NtDp is likely because the increase in deuteration of \Hthp outweighs the decrease in \Nt due to freeze-out.
Regarding the deuteration of \NtHp, we have observed a $X$(\NtDp)/$X$(\NtHp) profile spanning from an upper limit of 0.07 in the outer region to 
a maximum of 0.27$_{-0.15}^{+0.12}$ toward the center of L\,1498. 
However, this maximum value is moderate compared to values obtained in the other two starless cores, L\,183 \citep{Pagani07} and L\,1512 \citep{Lin20}. 
The authors found a maximum \NtHp deuteration of 0.70$\pm$0.12 in L\,183 and 0.34$_{-0.15}^{+0.24}$ in L\,1512.
It is worth noting that the formation and destruction of \NtDp and \DCOp follow the same chemical scheme in starless cores \citep{Pagani11}. 
Given that \NtDp and \DCOp are formed under the same conditions of \Hthp deuteration, 
the presence of \DCOp depletion suggests that the depletion of CO is greater than that of \Nt.

\subsection{CO and \Nt abundance profiles} \label{sec:discussion_mole_L1498}

Molecular nitrogen is the parent molecule reacting with the \Hthp isotopologues to form \NtHp and \NtDp, while CO is the parent molecule also reacting with \Hthp isotopologues to form \DCOp \citep{Pagani11}. In starless cores, \NtHp, \NtDp, \DCOp are formed via the above gas-phase paths. 
Therefore, as we presented in Sect.\,\ref{sec:analysis_chem_L1498}, $X$(\Nt) and $X$(CO) are free parameters for fitting the observed abundances of their daughter cations. 
The top row in Fig.\,\ref{fig:chem_model_L1498_CO_N2} shows the abundance profiles of \Nt, \twvCO, and \CetO.
Both \twvCO and \CetO have the same depletion factors of $\sim$20 compared with their maximum abundances of $1.0\times10^{-5}$ for \twvCO and $1.9\times10^{-8}$ for \CetO
since we assumed a constant ratio of $^{12}$C$^{16}$O/$^{12}$\CetO as 560 to derive the \CetO profile from \twvCO profile. 
By comparing with the standard \twvCO abundance of $\mbox{(1--2)}\times10^{-4}$ \citep{Pineda10}, \twvCO has a depletion factor of $\sim$200--400. For \CetO, the depletion factor is $\sim$190 with respect to its standard abundance of $1.7\times10^{-7}$ \citep{Frerking82}.
Conversely, the \Nt abundance does not decrease but even increases toward the core center (Fig. \ref{fig:chem_model_L1498_CO_N2}).
This accounts for the discrepancy between the depletion of \DCOp and the enhancement of \NtDp toward the core center.

In our chemical modeling, we find that the dust grain radius increases from 0.1~\micron in the outer region to 0.4~\micron in the core center. 
This grain growth slows down the depletion of CO and \Nt due to freeze-out onto dust grains at the core center by a factor of $\sim$3, as the depletion timescale is proportional to the dust grain radius \citep{Maret13}.
However, the freeze-out rates of CO and \Nt are suggested to be similar because of their identical mass and similar binding energies \citep{Bisschop06}. 
Also, the cosmic ray ionization is too weak in the dense core to compensate for the depletion of CO and \Nt.
The discrepancy between the \Nt and CO abundance profiles may suggest that the \Nt production from the atomic N 
is still active such that the \Nt production can compensate for the depletion of \Nt, similar to what \citet{Pagani12} found in the core center of L\,183.
In contrast, the CO production has reached a steady-state in the diffuse cloud region \citep{Oppenheimer75}, making the freeze-out effect the dominant mechanism regulating the CO abundance in the dense core region.

With the envelope tracers of CO and HCO$^{+}$,
\citet{Maret13} performed pseudo-time-dependent chemical modeling to study their depletion in L\,1498. 
They constrained their model by fitting the integrated intensity profiles and the central spectra of \CetO (1--0) and H$^{13}$CO$^+$ (1--0).
To reproduce their observations, 
they found a maximum grain radius of 0.15~\micron in the core center.
Our maximum grain radius of 0.4~\micron is greater than their value, likely because the four molecular cations (\NtHp, \NtDp, o-\HtDp, and \DCOp) we used are not heavily depleted and their emissions are optically thinner than the above envelope tracers. Therefore, our chemical modeling is more sensitive to the grain radius in the central region.
Despite this difference,
when they included such grain growth in their chemical modeling, 
they derived a central CO abundance of $\sim6\times10^{-7}$, which is consistent with our value of 5.0$_{-1.0}^{+26.6}\times10^{-7}$.
In addition, their chemical network includes nitrogen chemistry.
They found that a certain fraction of the nitrogen should be locked in \NHth ices to reduce the gas-phase abundances of \Nt, and atomic N. 
As a result, they obtained a \NtHp abundance of $4\times10^{-10}$, which is closer to $1.7\times10^{-10}$ derived by \citet{Tafalla04} (also see Sect.\,\ref{sec:discussion_ion_L1498} for the comparison of our \NtHp abundance profile with \citeauthor{Tafalla04}).
However, our central $X$(\NtHp) is even lower at $4.7\pm1.7\times10^{-11}$, and our central $X$(\Nt) at 4.0$_{-2.0}^{+2.3}\times10^{-6}$ is larger than \citet{Maret13}'s $2\times10^{-7}$. 
Therefore, more detailed modeling is necessary to understand the nitrogen chemistry in L\,1498.

\subsection{Core age of L\,1498} \label{sec:discussion_lifetime_L1498}

Figure\,\ref{fig:chem_model_L1498} shows that between 0.07 to 0.31 Ma,
the chemical solution of each layer appears at the observed abundances within the uncertainties.
One important characteristic of deuterium chemistry in starless cores is that the deuteration fractionation 
can serve as an effective chemical clock.
This is because it primarily depends on time and is influenced by a few initial parameters, including the initial ortho-to-para ratio of \Ht (OPR$_{\rm intial}$(\Ht)), the cosmic ray ionization rate ($\zeta$), and the average grain radius ($a_{\rm gr}$).
It is less affected by temperature and dynamical parameters such as the turbulent Mach number and the mass-to-magnetic flux ratio \citep{Pagani13, Kong16, Kortgen17}.
Therefore, given reasonable values/ranges of OPR$_{\rm intial}$(\Ht), $\zeta$, and $a_{\rm gr}$, 
the timescale of deuteration fractionation (i.e., the chemical timescale) provides a means to measure the age of a contracting core.

With pseudo-time-dependent chemical analysis, 
we have determined the chemical timescales of the deuteration in each layer in Sect.\,\ref{sec:analysis_chem_L1498}.
Since the physical structure is kept constant at the present state in the pseudo-time-dependent models during the evolution of the chemical composition, our chemical timescales do not include the physical evolution of the core.
Therefore, the determined chemical timescales at the fixed densities serve as a lower limit to the the core age.
Particularly, we take the chemical timescale of a less evolved outer layer, where $X$(\NtDp)/$X$(\NtHp) ratio and $X$(o-\HtDp) are measured, to represent the lower limit of the core age.
It follows that the lower limit on the core age of L\,1498 is $\sim$0.16~Ma determined by the fourth southwestern layer, which is the last layer with the \NtDp/\NtHp ratio measurement, and 
this lower limit is conservatively chosen at the beginning of the time range of this layer.
We would like to note that 
we use the $X$(\NtDp)/$X$(\NtHp) ratio and $X$(o-\HtDp) to determine the chemical timescale of a layer, while we use $X$(\DCOp) to derive the CO abundance.
Since the fourth southwestern layer has no o-\HtDp detection but only an upper limit on its abundance, 
the interpretation for this time limit is the time duration for reaching $X$(\NtDp)/$X$(\NtHp) under the assumption of $X$(o-\HtDp) at the maximum of its possible range, that is the fastest time of that layer. 

Although deuteration fractionation depends on OPR$_{\rm intial}$(\Ht), $\zeta$, and $a_{\rm gr}$, we assumed OPR$_{\rm intial}$(\Ht) to be the statistical ratio of 3. Additionally, we adopted $\zeta=3\times10^{-17}$~s$^{-1}$ from \citet{Maret13} and the typical $a_{\rm gr}$ of 0.1\micron for the the fourth southwestern layer (see Sect.\,\ref{sec:analysis_chem_L1498}).
Therefore, L\,1498 is presumably older than 0.16~Ma. 
If we assume another OPR$_{\rm intial}$(\Ht) of 0.5 as measured in the diffuse medium \citep{Crabtree11}, we find another, but similar, age lower limit of 0.13~Ma.
In the fast collapse model, the contraction timescale of the core can be characterized by the free-fall time, $t_{\rm ff}$.
\begin{equation}
t_{\rm ff}=\sqrt{\frac{3\pi}{32G\rho}},
\end{equation}
where $G$ is the gravitational constant and
$\rho$ represents the mass density, which can be expressed as $\rho=\mu_{\rm H_2}m_{\rm H}n_{\rm H_2}$, where $m_{\rm H}$ is the mass of atomic hydrogen and $\mu_{H_2}=2.8$ denotes the mean molecular weight per \Ht.
Considering that dense starless cores are typically contracted from
lower density gas at $n_{\rm H_2}=10^4$~cm$^{-3}$, $t_{\rm ff}=0.31$~Ma, which is compatible with the lower limit of the core age of L\,1498 at 0.16~Ma.
On the other hand, the core contraction timescale in the slow collapse model can be associated with the ambipolar diffusion timescale in the presence of the magnetic field support. The ambipolar diffusion timescale can be longer than a free-fall time by an order of magnitude \citep{Shu87, Tassis04, Mouschovias06, McKee07}.
Although a lower limit of the core age at 0.16~Ma does not conclusively rule out the possibility of a much longer actual core age,
such a long contraction timescale does not seem to be favored by our findings. 
Particularly, significant inward motions have been detected toward the L\,1498 envelope via CS lines \citep{Lee01, Lee11} and HCN lines \citep{Magalhaes18}.
Therefore, our result suggests that the contraction of L\,1498 follows the fast collapse model, indicating that L\,1498 likely formed rapidly.

\citet{Maret13} and \citet{Jimenez-Serra21} also derived the chemical timescales of L\,1498 but with different molecules.
\citet{Maret13} derived a chemical timescale of L\,1498 with a range of 0.3--0.5 Ma based on the CO freeze-out modeling with the \CetO (1--0) line,
while \citet{Jimenez-Serra21} derived a chemical timescale as 0.09 Ma from their COMs observation toward the methanol emission peak at the outer layers of L\,1498.
Because both studies employed the pseudo-time-dependent chemical analysis, their chemical timescales are also lower limits on the age of L\,1498.
It is noteworthy that although the timescale derived by \citeauthor{Maret13} is longer than our deuteration timescale, using their value as the lower limit of the core age would still favor the fast collapse model. 
In fact, $X$(CO) at the core center derived from our deuteration chemical modeling aligns with their derived CO value (see Sect\,\ref{sec:discussion_mole_L1498}).
However, assessing CO depletion toward highly embedded and more evolved cores can be challenging due to the contamination along the sightline and the severe CO depletion. 
In such cases, the deuteration fraction is easier to measure and can provide valuable insights into the chemical evolutionary history of the core contraction.
Therefore, it is demonstrated that different chemical clocks provide a comprehensive point of view on the core ages.
On the other hand, in the chemical modeling of COMs, the surrounding environment could be more important than chemical evolutionary effects \citep{Lattanzi20}. Then COMs may not act as a chemical clock but as an indicator of the environmental conditions.

In this work, we discuss the low-mass core formation in 
the context of contrasting fast and slow collapse models of relatively isolated entities in the parent clouds. 
However, we note that 
the interactions between starless cores and their surrounding clumps/filaments may lead to more dynamical scenarios,
such as fragmentation into smaller entities and mass accumulation/accretion from parent filamentary structures, suggesting a hierarchical nature \citep[e.g.,][]{Vazquez-Semadeni19}. 
This complexity potentially aligns with scenarios favoring fast collapse.
Recently, \citet{Bovino21} 
performed a 
3D turbulent MHD simulation coupled with deuterium
chemistry to study a collapsing filament, fragmenting into dense cores.
They proposed the o-\HtDp to p-\DtHp ratio as an alternative chemical clock to the \NtDp/\NtHp ratio for tracing OPR(\Ht).
Their APEX o-\HtDp (\oHtDpgrd) and p-\DtHp (\pDtHpgrd) observations of six starless cores in the Ophiuchus region demonstrate that fast collapse models align well with the observed o-\HtDp/p-\DtHp ratios
with core ages at $\sim$0.05--0.2~Ma, indicating 
that these Ophiuchus cores can form more rapidly.

\section{Conclusions}\label{sec:conclusion_L1498}

We present the first o-\HtDp (\oHtDpgrd) detection toward L\,1498.
We carried out a non-LTE radiative transfer calculation
with an asymmetric onion-like model to 
reproduce the emission line of L\,1498. 
We obtained 3D asymmetric profiles of 
the physical parameters and chemical abundances of \NtHp, \NtDp, \DCOp, and o-\HtDp.
Then, we used a time-dependent chemical model adopting a deuterium chemical network to estimate the lower limit on the core age of L\,1498 by fitting the chemical solutions
to the observed abundances. 
We summarize our results as follows:
\begin{enumerate}
\item We derived the central molecular hydrogen density of $1.6_{-0.3}^{+3.0}\times10^{5}$~cm$^{-3}$ 
and the central kinetic temperature of 7.5$_{-0.5}^{+0.7}$~K from \NtHp observations, resulting in a peak $N_{\rm H_2}$ of $4.5\times10^{22}$~cm$^{-2}$ toward the L\,1498 core.

\item We found that the depletion factors of 5.6$_{-4.0}^{+2.1}$ for \NtHp and 17$\pm$11 for \DCOp.
The deuterium fractionation of \NtHp is enhanced from an upper limit 
of $\leq 0.07$ at large radii to 0.27$_{-0.15}^{+0.12}$ in the center.

\item The opposite behaviors of the \NtDp enhancement and \DCOp depletion toward the core center suggest that the depletion of CO is greater than that of \Nt.
Our results show that \twvCO and \CetO have a depletion factor of $\sim$20 between internal and external layers in the core. Comparing their minimum abundance to their standard (literature) abundance, both isotopologues have 
comparable depletion factors of about 200.

\item We derive a lower limit on the core age of L\,1498 as 0.16~Ma which is comparable to the typical free-fall time of 0.31~Ma, assuming that starless cores contract from gas with a typical density of $n_{\rm H_2}=10^4$~cm$^{-3}$.
Our result suggests that the contraction of L\,1498 follows the fast collapse model, indicating that L\,1498 likely formed rapidly.

\end{enumerate}
In summary, our results show a self-consistent description of density, temperature, and molecular abundances in a starless core.
We have expanded our investigation to include a survey of additional low-mass cores.
Through the detailed modeling of our samples, 
comparing them will shed light on discriminating factors in their evolution. We would offer a comprehensive understanding of core formation mechanisms by assessing their ages and dynamics.

\begin{acknowledgements}
The authors would like to warmly thank Mario Tafalla (Instituto Geografico Nacional Observatorio Astron\'{o}mico Nacional, Spain) for providing his 1.2\,mm MAMBO and molecular line maps, 
S\'{e}bastien Maret (Institut de Plan\'{e}tologie et d'Astrophysique de Grenoble, France) for providing his molecular line maps,
Sheng-Yuan Liu and I-Ta Hsieh (ASIAA, Taiwan) for discussing the details of the SPARX code, 
David T. Frayer (GBT) for discussing the details of beam efficiencies,
and
Nawfel Bouflous and Patrick Hudelot (TERAPIX data center, IAP, Paris, France) for their help in preparing the CFHT/WIRCAM observation 
scenario and for performing the data reduction.
S.J.L. acknowledges the grants from 
the National Science and Technology Council (NSTC) of
Taiwan 111-2124-M-001-005 and 112-2124-M-001-014.
S.J.L. and S.P.L. acknowledge the grants from 
NSTC of
Taiwan 109-2112-M-007-010-MY3, 112-2112-M-007-011.
This work used high-performance computing facilities operated by the
Center for Informatics and Computation in Astronomy (CICA) at NTHU. This equipment was funded by the Ministry of
Education of Taiwan, the Ministry of Science and Technology of Taiwan,
and NTHU.
This work was supported by the Programme National ``Physique et Chimie du 
Milieu Interstellaire'' (PCMI) of INSU/CNRS with INC/INP co-funded by CEA 
and CNES and by Action F\'ed\'eratrice Astrochimie de l'Observatoire de Paris.
This work is based in part on observations carried out under project 
numbers 152-13, 039-14, and 112-15 with the IRAM 30-m telescope. 
IRAM is supported by INSU/CNRS (France), MPG (Germany) and IGN (Spain). 
The JCMT data were collected during program M15BI046 (HARPS, and M17BP043 but data were lost due to an incorrect tuning).
The JCMT is operated by the East Asian Observatory on behalf of The National Astronomical Observatory of Japan; Academia Sinica Institute of Astronomy and Astrophysics; the Korea Astronomy and Space Science Institute; the National Astronomical Research Institute of Thailand; Center for Astronomical Mega-Science (as well as the National Key R\&D Program of China with No. 2017YFA0402700). Additional funding support is provided by the Science and Technology Facilities Council of the United Kingdom and participating universities and organizations in the United Kingdom and Canada.
The CFHT is operated by the National Research Council (NRC) of Canada, the Institute National des Sciences de l'Univers of the Centre National de la Recherche Scientifique of France, and the University of Hawaii. 
The observations at JCMT and CFHT were performed with care and respect from the summit of Maunakea.
The authors wish to recognize and acknowledge the very significant cultural role and reverence that the summit of Maunakea has always had within the indigenous Hawaiian community.  
We are most fortunate to have the opportunity to conduct observations from this mountain.
This research made use of the Aladin interface, the SIMBAD database, operated
at CDS, Strasbourg, France, and the VizieR catalog access tool, CDS, Strasbourg, France. 
This research has made use of the NASA/IPAC Infrared Science Archive, which is 
operated by the Jet Propulsion Laboratory, California Institute of Technology, 
under contract with the National Aeronautics and Space Administration. 
This work is based in part on observations made with the Spitzer Space Telescope, 
which is operated by the Jet Propulsion Laboratory, California Institute of Technology 
under a contract with NASA. 
This research used the facilities of the Canadian Astronomy Data Centre operated by 
the National Research Council of Canada with the support of the Canadian Space Agency.
Green Bank Observatory is a facility of the National Science Foundation and is 
operated by Associated Universities, Inc. 
\end{acknowledgements}

\bibliographystyle{aa}
\bibliography{ref}

\begin{appendix}

\section{Antenna temperature scale and the GBT beam efficiencies}
\label{app:Ta}

The antenna temperature ($T_{\rm A}^*$) is the forward beam brightness temperature corrected for atmosphere, ohmic, and rearward losses \citep{Kutner81}.
Given the source brightness distribution of $T_{\rm b}(\theta, \phi)$ from our radiative transfer models, the modeled intensity in the $T_{\rm A}^*$ scale resulting from the telescope coupling is
\begin{equation}
T_{\rm A}^* = \frac{P_{\rm 2\pi}\ast T_{\rm b}}{\Omega_{\rm 2\pi}} 
= \frac{\Omega_{\rm MB}}{\Omega_{\rm 2\pi}} \frac{P_{\rm MB}\ast T_{\rm b}}{\Omega_{\rm MB}} + 
{\textstyle \sum_i} \frac{\Omega_{{\rm EB}, i}}{\Omega_{\rm 2\pi}} \frac{P_{{\rm EB}, i}\ast T_{\rm b}}{\Omega_{{\rm EB}, i}},
\end{equation}
where ${P_{\rm 2\pi}=P_{\rm 2\pi}(\theta, \phi)}$ is the telescope forward beam pattern consisting of the main beam and the $i$th error beam components (i.e., ${P_{\rm 2\pi}=P_{\rm MB}+\sum_i P_{{\rm EB}, i}}$), and
$\Omega_{\rm 2\pi}$ ($\Omega_{\rm MB}$, and $\Omega_{{\rm EB}, i}$) is the integral of the beam pattern over the forward hemisphere (the main beam, and the $i$th error beam components).
Since the error beam patterns are approximated by 2D Gaussian functions in practice, the error beam sizes, $\theta_{{\rm EB}, i}$, are defined as the HPBWs \citep{Bensch01, Greve98}.
Here, we denote the ``$\ast$'' symbol as the convolution operator
\footnote{$P\ast T_{\rm b}=\iint_{\rm 2\pi} P(\theta, \phi)T_{\rm b}(\theta - \theta_0, \phi - \phi_0)\,d\Omega$ with the telescope pointing toward the direction of $(\theta_0, \phi_0)$.}.

Then ${T_{\rm MB,C}=P_{\rm MB}\ast T_{\rm b}/\Omega_{\rm MB}}$ is defined as the corrected main beam temperature\footnote{\citet{Bensch01} introduced ${T_{\rm MB,C}=P_{\rm MB}\ast T_{\rm b}/\Omega_{\rm MB}}$ as the ``corrected main beam temperature'' to exclude the error beam contributions ($P_{\rm EB}$)
in contrast to the common main beam temperature definition of ${T_{\rm MB}=P\ast T_{\rm b}/\Omega_{\rm MB}}$. We can see that $T_{\rm MB}$ will be over-corrected (${T_{\rm MB,C}<T_{\rm MB}}$) with non-negligible $P_{\rm EB}$.}, 
while ${T_{{\rm EB}, i}=P_{{\rm EB}, i}\ast T_{\rm b}/\Omega_{{\rm EB}, i}}$ is defined as the $i$th error beam temperature. 
The corresponding main beam and the $i$th error beam efficiencies are defined by
${\eta_{\rm MB}=\Omega_{\rm MB}/\Omega_{\rm 2\pi}}$ and 
${\eta_{{\rm EB}, i}=\Omega_{{\rm EB}, i}/\Omega_{\rm 2\pi}}$, respectively.
Hence the expression of $T_{\rm A}^*$ can be written as
\begin{equation}
T_{\rm A}^* = \eta_{\rm MB} T_{\rm MB, C}+ {\textstyle \sum_i} \eta_{{\rm EB}, i} T_{{\rm EB}, i}.
\end{equation}

Table\,\ref{tab:gbt_eff} shows the GBT beam pattern measurements at 77~GHz,
which we have adopted in our modeling of \DCOp (1--0) and \NtDp (1--0) spectra.
As the GBT technical report\footnote{https://www.gb.nrao.edu/scienceDocs/GBT\_302.pdf} provides detailed beam pattern measurements only at 86~GHz,
we derived the corresponding measurements at 77~GHz based on beam efficiencies provided by 
David T. Frayer (private communication).
At 77~GHz, the main beam efficiency is ${\eta_{\rm MB}=0.51}$, while
the forward scattering and spillover efficiency ($\eta_{\rm fss}$)
the forward efficiency ($\eta_{\rm l}$) are comparable to those at 86~GHz (i.e., $\eta_{\rm fss}=0.97$ and $\eta_{\rm l}=0.98$).
We assume two error beams with equal $\eta_{\rm EB}$ values, similar to the 86~GHz error beam pattern. Therefore, we obtain $\eta_{\rm EB}=(\eta_{\rm fss}-\eta_{\rm MB})/2=0.23$. Furthermore, the error beam sizes are scaled from those at 86~GHz by that $\theta_{\rm EB}$ is inversely proportional to the frequency.

\begin{table}[h!]
    \centering
    \caption{GBT beam pattern measurements at 77~GHz}
    \begin{tabular}{lcc}
        \hline\hline
        Beam component & $\eta_{\rm MB}$ or $\eta_{\rm EB}$ & $\theta_{\rm MB}$ or $\theta_{\rm EB}$\\
        \hline
        Main beam & 0.51 & 11\arcsec \\
        First error beam & 0.23 & 61\arcsec \\
        Second error beam & 0.23 & 2100\arcsec \\
        \hline
    \end{tabular}
    \label{tab:gbt_eff}
\end{table}

\section{Ortho-\HtDp (\oHtDpgrd) and \NtHp (1--0) spectral line observations}\label{app:all_spectra_L1498}

We present our full o-\HtDp (\oHtDpgrd) spectral line observations and the \NtHp (1--0) spectral line observations obtained by \citet{Tafalla04}, comparing with our best-fit asymmetric onion-shell model.

\begin{figure*}[h!]
	\centering
	\includegraphics[width=0.95\textwidth]{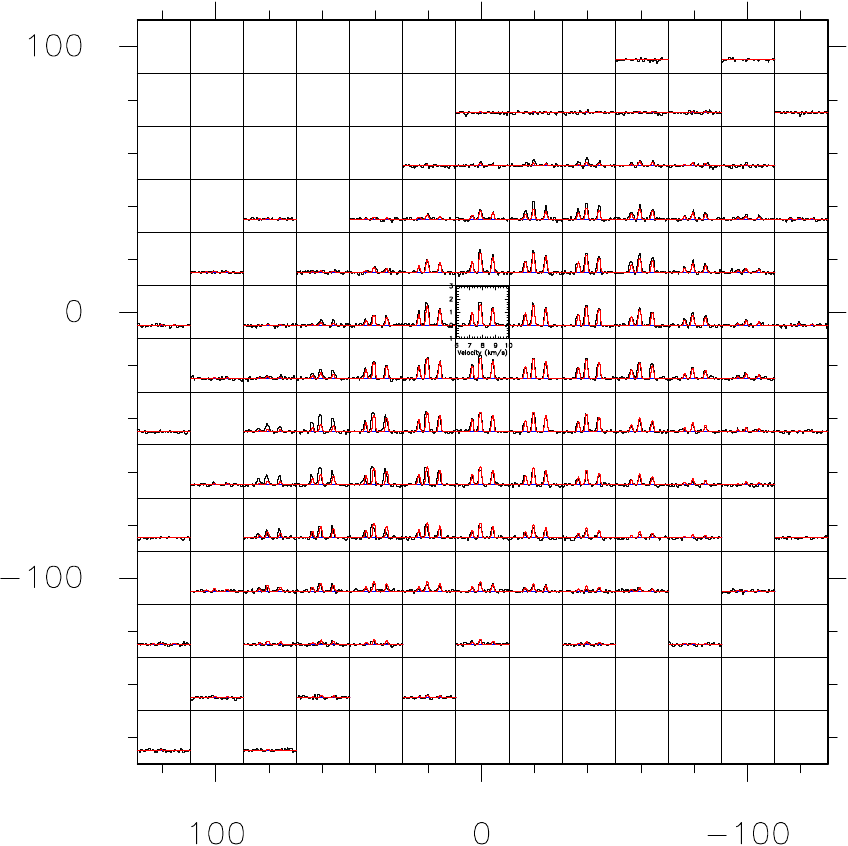}
	\caption{
    \NtHp (1--0) spectra from \citet{Tafalla04}.
    The $x$- and $y$-axes of the grid are the RA and Dec offsets with respect to the center of L\,1498.
    Each cell shows the observational spectra as black, our modeled spectra as red, and the baselines as blue. The dimension of $T_{\rm A}^*$ and $V_{\rm LSR}$ at each cell are $-$1$\sim$3~K and 6$\sim$10~km~s$^{-1}$, respectively, denoted in the central cell. Only the central triplet is shown for clarity.
    \label{fig:n2hp10_spectra_L1498}
    }
\end{figure*}

\begin{figure*}[h!]
	\centering
	\includegraphics[width=0.95\textwidth]{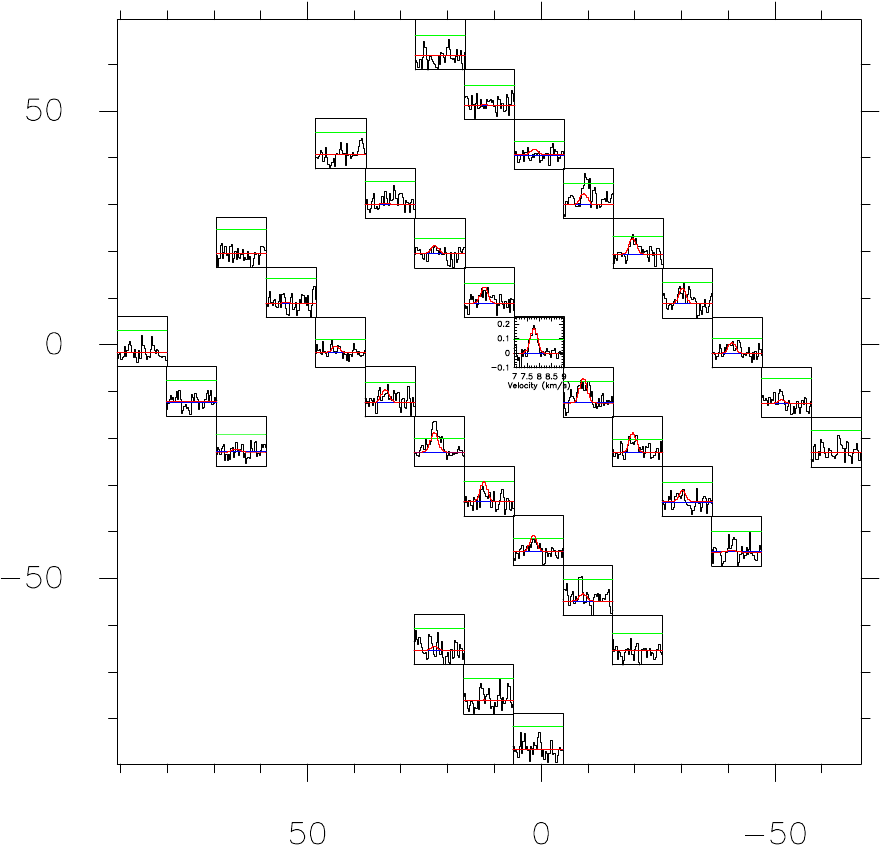}
	\caption{
    o-\HtDp (\oHtDpgrd) spectra.
    The $x$- and $y$-axes of the grid are the RA and Dec offsets with respect to the center of L\,1498.
    Each cell shows the observational spectra as black, the modeled spectra as red, and the baselines as blue. The 3$\sigma$ noise levels are shown in green horizontal lines.
    The dimensions of $T_{\rm A}^*$ and $V_{\rm LSR}$ at each cell are $-$0.1$\sim$0.25~K and 7$\sim$9~km~s$^{-1}$, respectively, denoted in the cell closest to the core center, (0.5\arcsec, 0.5\arcsec).
    \label{fig:h2dp_spectra_L1498}
    }
\end{figure*}

Figures 
\ref{fig:n2hp10_spectra_L1498}  
and 
\ref{fig:h2dp_spectra_L1498}
show all the single pointing observations in black lines and models of 
\NtHp (1--0) 
and 
o-\HtDp (\oHtDpgrd) 
in red lines, respectively. In addition, the 3$\sigma$ noise levels are denoted by green horizontal lines in Fig.\,\ref{fig:h2dp_spectra_L1498}.
The models are reproduced with our asymmetric onion-like model (Sect.\,\ref{sec:analysis_L1498} and Table \ref{tab:rad_model_L1498}).


\section{\CetO spectral line observations}\label{app:all_spectra_L1498_c18o}
We present the \CetO (1--0) spectral line observations obtained by \citet{Tafalla04} along the main cut across our model center.
As the discussion in Sect. \ref{sec:discussion_den_L1498}, Figure \ref{fig:Av_fitting_L1498} indicates that the L\,1498 core is embedded in an envelope consisted of a layer with \Av $\approx$3~mag.
The critical density at $T_{\rm kin}=$10~K is $1.9\times10^3$~cm$^{-3}$ for the $J$=1--0 line and $8.4\times10^3$~cm$^{-3}$ for the $J$=2--1 line.
Our \CetO (2--1) spectral line observations shown in the bottom row in Fig.\,\ref{fig:maincut_spectra_L1498} suggest that the envelope does not emit pronounced  \CetO (2--1) line emission compared to the core region. Therefore, the number density of the envelope should be much lower than the critical density of the $J$=2--1 line but slightly larger than that of the $J$=1--0 line.

\begin{figure*}[h!]
	\centering
	\includegraphics[width=\textwidth]{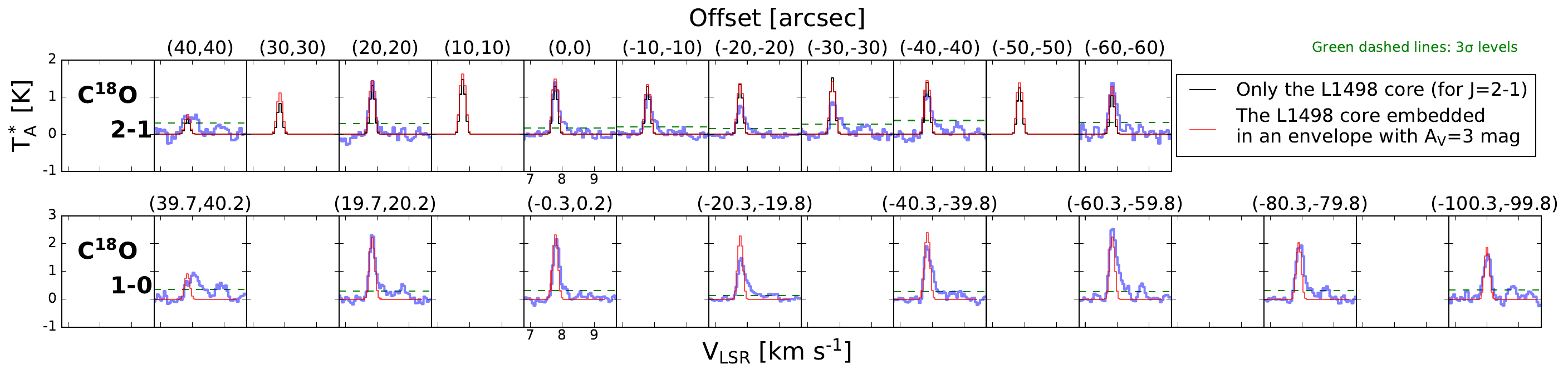}
	\caption{
    \CetO (1--0) and (2--1) spectra along the main cut.
    The blue spectra show the observational data including the archival $J$=1--0 spectra from  \citet{Tafalla04}
    and our $J$=2--1 spectra.
    The black spectra show the modeled $J$=2--1 spectra from Fig.\,\ref{fig:maincut_spectra_L1498}.
    The red spectra show the modeled spectra of the asymmetric onion-like model embedded in an envelope that is a layer with \Av=3~mag.
    Each column corresponds to different offsets from the center of L\,1498. 
    The green dashed lines indicate the 3$\sigma$ noise level.}
    \label{fig:c18o10_spectra_L1498}
\end{figure*}

We find that $n_{\rm H_2}=2\times10^3$~cm$^{-3}$ can reproduce the \CetO (1--0) spectra without enhancing the \CetO (2--1) spectra. 
Figure \ref{fig:c18o10_spectra_L1498} shows the single pointing observations in blue thick lines.
The red \CetO spectrum models are based on radiative transfer calculations performed with the asymmetric onion model (the L\,1498 core) embedded in the envelope, while 
the black \CetO (2--1) spectrum models are performed with only the asymmetric onion model of the L\,1498 core.
In addition, the 3$\sigma$ noise levels are denoted by green dashed horizontal lines.

With ${N_{\rm H_2}/A_{\rm V}=9.4\times10^{20}}$~cm$^{-2}$~mag$^{-1}$ \citep[\Rv= 3.1;][]{Bohlin78}, \Av=3~mag is equivalent to ${N_{\rm H_2}=2.8\times10^{21}}$~cm$^{-2}$, and the thickness of the envelope is 0.46~pc (94~kau) along the line of sight with $n_{\rm H_2}$ of $2\times10^3$~cm$^{-3}$.
For $T_{\rm kin}$, we adopt $T_{\rm kin}=10$~K or the best-fit temperature from the outermost onion-like layer if it is larger than 10~K (Table \ref{tab:rad_model_L1498}). 
Thus they are $T_{\rm kin}=10$~K in the southwestern envelope, and 
$T_{\rm kin}=12$~K the northeastern envelope.
For $X$(\CetO), we adopt the standard abundance of $1.7\times 10^{-7}$ \citep{Frerking82} except that the \CetO abundance in the most northeastern region in the northeastern envelope is decreased to the same abundance in the fifth northeastern onion-like layer as $1.9\times10^{-8}$ in order to match the spectrum at ($\Delta\mbox{RA}=40$\arcsec, $\Delta\mbox{Dec}=40$\arcsec). This lower \CetO abundance may be due to the selective photodissociation at the northeastern cloud edge.

\section{Asymmetric onion-like model}\label{app:asym_model}

\begin{figure*}
	\centering
	\includegraphics[scale=0.8]{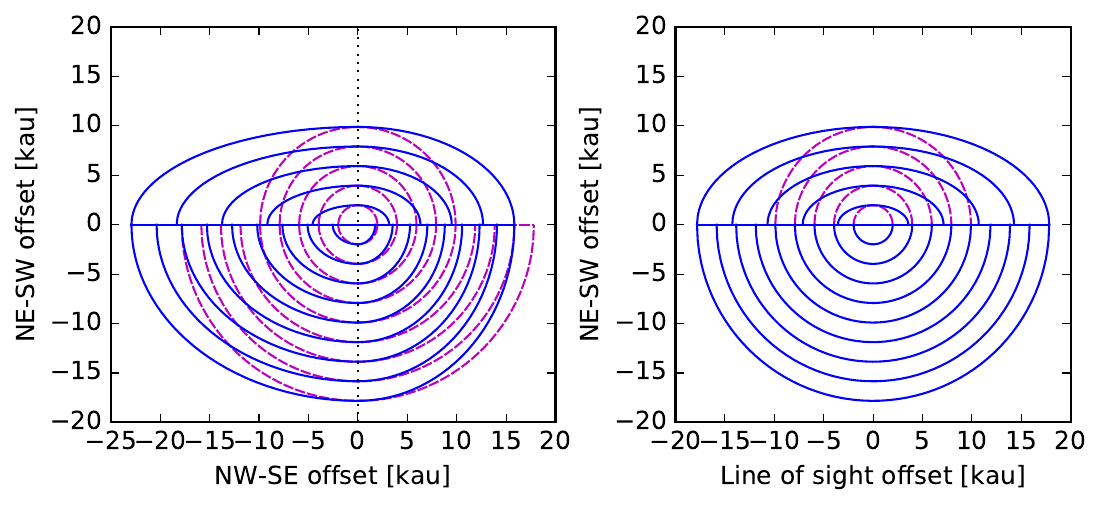}
	\caption{
    Comparison between the asymmetric and spherical onion-like models.
    The asymmetric layers are shown in the blue solid curves, while 
    the spherical layers are shown in the magenta dashed curves. 
    The left panel shows the cross-section of the models across the center and within the sky plane.
    The right panel shows the cross-section along the main cut (NE-SW direction) and the line of sight, where both asymmetric and spherical models are identical on the NE side.}
    \label{fig:sym2asym}
\end{figure*}

Figure \ref{fig:sym2asym} shows the comparison between the spherically dissymmetric and asymmetric onion-like models. 
The layer width in the spherical model is chosen to be $10\sqrt{2}$ arcsec ($\approx 14.1$\arcsec and 1980~au at the distance of 140~pc), which is the same as the spacing of our IRAM 30-m/GBT pointing observations (Fig.\,\ref{fig:pointing_L1498}).
We note that the spherically dissymmetric model is comprised of two spherically symmetric models for the northeast and southwest sides with five and nine layers, respectively.
We took the emission morphologies of 850~\micron continuum (Fig.\,\ref{fig:maps_L1498}h) and the integrated intensity of \NtHp (1--0)  (Fig.\,\ref{fig:maps_L1498}i) as the reference. 
Since the southwest side of L\,1498 is rounder than the other side, we intended to keep the southwestern geometric shape at a cross-section along the main cut and sightline (i.e., a semi-circle) the same in both symmetric and asymmetric models.
We then modified the southwest side of the spherical model to become the asymmetric model by shrinking the west quadrant by a factor of 8/9, and extending the south quadrant by a factor of 9/7 along the NW--SE direction.
On the northeast side, the semi-sphere extends along both the NW--SE direction and the sightline direction in order to match with the southwest side.

We choose the NW--SE axis as the $x$-axis (the positive direction is PA $= -45^\circ$), 
the NE--SW axis as the $y$-axis (the positive direction is PA $= +45^\circ$),
and the line-of-sight axis as the $z$-axis.
Supposing the radius of a layer boundary in the spherical model is $r_{\rm sph}$, then 
the corresponding layer boundary in the asymmetric model,
$r_{\rm asym}=(x,y,z)$,
is defined in the four quadrants on the plane of sky by the following equations.

In the north quadrant ($x>0, y>0$),
\begin{equation}
1 = \frac{x^2}{\big(\frac{8}{9}\times\frac{9}{5}r_{\rm sph}\big)^2}
+ \frac{y^2}{r_{\rm sph}^2}
+ \frac{z^2}{\big(\frac{9}{5}r_{\rm sph}\big)^2}.
\end{equation}\label{equ:asym_1}

In the east quadrant ($x<0, y>0$),
\begin{equation}
1 = \frac{x^2}{\big(\frac{9}{7}\times\frac{9}{5}r_{\rm sph}\big)^2}
+ \frac{y^2}{r_{\rm sph}^2}
+ \frac{z^2}{\big(\frac{9}{5}r_{\rm sph}\big)^2}.
\end{equation}

In the south quadrant ($x<0, y<0$),
\begin{equation}
1 = \frac{x^2}{\big(\frac{9}{7}r_{\rm sph}\big)^2}
+ \frac{y^2}{r_{\rm sph}^2}
+ \frac{z^2}{r_{\rm sph}^2}.
\end{equation}

In the west quadrant ($x>0, y<0$),
\begin{equation}
1 = \frac{x^2}{\big(\frac{8}{9}r_{\rm sph}\big)^2}
+ \frac{y^2}{r_{\rm sph}^2}
+ \frac{z^2}{r_{\rm sph}^2}.
\end{equation}\label{equ:asym_4}

\section{Best-fit physical and abundance profiles}\label{app:rad_model_table}

We present the best-fit quantities with their error ranges for each layer in our asymmetric onion-shell model from Fig.\,\ref{fig:rad_model_L1498} and Fig.\,\ref{fig:chem_model_L1498_CO_N2} in Table\,\ref{tab:rad_model_L1498}.

\begin{sidewaystable*}[hb]
\caption{Onion-shell model parameters.}
\centering
\begin{tabular}{rcrccrccccc}
\hline\hline
$r$        & $n_{\rm H_2}$& $T_{\rm kin}$ & $X$(\NtHp)                 & $X$(\NtDp)   & $D_{\rm N_2H^+}$$^{\tablefootmark{a}}$ & $X$(\DCOp)                 & $X$(o-\HtDp) & $X$(CO)      & $X$(C$^{18}$O) & $X$(N$_2$) \\ 
(kau)      & (cm$^{-3}$)  & (K)           &                            &              &                                        &                            &              &              &                &   \\
\hline
\noalign{\smallskip}
\multicolumn{11}{c}{Southwest side}\\
\noalign{\smallskip}
\hline
\noalign{\smallskip}
0.00--1.98   & 1.6$_{-0.3}^{+3.0}$(5)$^{\tablefootmark{b}}$ 
                                    & 7.5$_{-0.5}^{+0.7}$ & 4.7$_{-1.7}^{+1.7}$(-11) & 1.3$_{-0.5}^{+0.3}$(-11) & 0.27$_{-0.15}^{+0.12}$ & 5.2$_{-2.4}^{+3.0}$(-12) & 2.1$_{-0.9}^{+2.0}$(-10) & 5.0$_{-1.0}^{+26.6}$(-7) & 8.9(-10) & 4.0$_{-2.0}^{+2.3}$(-6) \\
\noalign{\smallskip}
1.98--3.96   & 1.5$_{-0.2}^{+3.0}$(5) & 7.5$_{-0.4}^{+0.8}$  & 8.2$_{-5.3}^{+0.8}$(-11) & 1.3$_{-0.5}^{+0.3}$(-11) & 0.16$_{-0.12}^{+0.04}$ & 6.4$_{-3.1}^{+2.2}$(-12) & 2.1$_{-0.8}^{+1.8}$(-10) & 5.0$_{-3.4}^{+10.8}$(-7) & 8.9(-10) & 1.0$_{-0.5}^{+0.1}$(-6) \\
\noalign{\smallskip}
3.96--5.93   & 1.3$_{-0.2}^{+3.7}$(5) & 7.5$_{-0.5}^{+0.8}$  & 1.0$_{-0.9}^{+0.0}$(-10) & 1.2$_{-0.6}^{+0.2}$(-11) & 0.11$_{-0.11}^{+0.02}$ & 1.1$_{-0.4}^{+0.5}$(-11) & 2.1$_{-1.0}^{+2.3}$(-10) & 7.9$_{-6.4}^{+92.1}$(-7) & 1.4(-9) & 5.0$_{-2.5}^{+2.9}$(-7) \\
\noalign{\smallskip}
5.93--7.92   & 9.2$_{-1.6}^{+18.0}$(4) & 7.5$_{-0.5}^{+0.7}$  & 1.5$_{-0.9}^{+0.1}$(-10) & 1.1$_{-0.5}^{+0.3}$(-11) & 0.07$_{-0.05}^{+0.02}$ & 5.0$_{-2.3}^{+1.4}$(-11) & <2.1(-10)                   & 6.3$_{-0.0}^{+33.5}$(-6) & 1.1(-8) & 1.0$_{-0.5}^{+2.2}$(-6) \\
\noalign{\smallskip}
7.92--9.90   & 6.8$_{-1.5}^{+17.2}$(4) & 7.5$_{-0.5}^{+0.7}$  & 2.6$_{-1.6}^{+0.3}$(-10)  & <1.1(-11)                  & <0.04                  & 8.9$_{-4.1}^{+2.3}$(-11) & --                         & 1.0$_{-0.4}^{+19.0}$(-5) & 1.9(-8) & 2.0$_{-1.0}^{+37.8}$(-6) \\
\noalign{\smallskip}
9.90--11.88  & 5.4$_{-0.8}^{+12.3}$(4) & 7.5$_{-0.6}^{+0.6}$ & 2.1$_{-0.8}^{+0.4}$(-10) & <8.5(-12)                  & <0.04                  & 5.3$_{-2.1}^{+2.1}$(-11) & --                          & 1.0$_{-0.6}^{+19.0}$(-5) & 1.9(-8) & 1.6$_{-0.6}^{+30.0}$(-6) \\
\noalign{\smallskip}
11.88--13.86 & 4.4$_{-1.1}^{+5.8}$(4) & 7.5$_{-0.7}^{+0.4}$ & 1.6$_{-0.7}^{+0.3}$(-10) & <6.5(-12)                  & <0.04                  & 3.0$_{-1.8}^{+0.8}$(-11) & --                          & 1.0$_{-0.8}^{+19.0}$(-5) & 1.9(-8) & 1.3$_{-0.6}^{+14.6}$(-6) \\
\noalign{\smallskip}
13.86--15.84 & 3.6$_{-0.9}^{+3.6}$(4) & 8.0$_{-0.8}^{+0.6}$ & 6.3$_{-2.5}^{+1.9}$(-11) & <2.5(-12)                  & <0.04                  & 1.2$_{-0.7}^{+0.3}$(-11) & --                          & 1.0$_{-0.9}^{+19.0}$(-5) & 1.9(-8) & 6.3$_{-3.1}^{+56.8}$(-7)\\
\noalign{\smallskip}
15.84--17.82 & 3.0$_{-0.7}^{+2.6}$(4) & 8.0$_{-0.8}^{+0.7}$  & 6.3$_{-1.5}^{+5.3}$(-11) & <2.5(-12)                  & <0.04                  & <1.2(-11)  & --                          & -- & 1.9(-8) & -- \\
\noalign{\smallskip}
\hline
\noalign{\smallskip}
\multicolumn{11}{c}{Northeast side}\\
\noalign{\smallskip}
\hline
\noalign{\smallskip}
0.00--1.98 & 1.6$_{-0.4}^{+2.8}$(5) & 7.5$_{-0.7}^{+0.6}$    & 4.7$_{-2.8}^{+0.2}$(-11) & 1.3$_{-0.4}^{+0.6}$(-11) & 0.27$_{-0.19}^{+0.13}$  & 5.2$_{-0.3}^{+6.7}$(-12) & 2.1$_{-0.4}^{+1.0}$(-10) & 7.9$_{-0.0}^{+71.5}$(-7) & 1.4(-9) & 4.0$_{-2.0}^{+1.0}$(-6) \\
\noalign{\smallskip}
1.98--3.96 & 7.2$_{-1.7}^{+12.9}$(4) & 8.5$_{-0.5}^{+0.7}$   & 4.0$_{-2.6}^{+0.1}$(-10) & 2.7$_{-0.8}^{+1.2}$(-11) & 0.07$_{-0.05}^{+0.03}$  & 7.6$_{-4.5}^{+1.1}$(-11) & <2.1(-10)                & 1.0$_{-0.5}^{+9.0}$(-5) & 1.8(-8) & 1.0$_{-0.6}^{+5.3}$(-5) \\
\noalign{\smallskip}
3.96--5.93 & 4.6$_{-0.8}^{+7.7}$(4) & 9.0$_{-0.7}^{+0.6}$    & 6.0$_{-2.8}^{+1.2}$(-11) & <3.6(-12)                & <0.06                   & 2.0$_{-0.6}^{+0.9}$(-11) & <2.1(-10)                & 1.0$_{-0.0}^{+19.0}$(-5) & 1.9(-8) & 4.0$_{-1.5}^{+59.1}$(-7) \\
\noalign{\smallskip}
5.93--7.92 & 2.4$_{-0.5}^{+2.3}$(4) & 10.0$_{-0.9}^{+0.4}$   & 2.9$_{-0.9}^{+2.0}$(-11) & <8.6(-13)                & <0.03                   & 1.3$_{-0.6}^{+0.3}$(-11) & --                       & 1.0$_{-0.7}^{+19.0}$(-5) & 1.9(-8) & 4.0$_{-0.8}^{+21.1}$(-7) \\
\noalign{\smallskip}
7.92--9.90 & 1.0$_{-0.3}^{+0.7}$(4) & 12.0$_{-1.7}^{+1.7}$   & 2.9$_{-0.6}^{+4.5}$(-11) & <8.6(-13)                & <0.03                   & <4.0(-12) & --                           & -- & 1.9(-8) & -- \\
\noalign{\smallskip}
\hline
\end{tabular}
\tablefoot{
\tablefoottext{a}{The deuteration ratio of \NtHp, $X$(\NtDp)/$X$(\NtHp).}
\tablefoottext{b}{It reads $1.6_{-0.3}^{+3.0}\times10^5$.}
}
\label{tab:rad_model_L1498}
\end{sidewaystable*}

\end{appendix}

\end{document}